\documentclass[submission,Phys]{SciPost}
\usepackage{graphicx,amsmath}
\usepackage{booktabs}
\usepackage{amssymb}
\usepackage{amsthm}
\usepackage{hyperref}
\usepackage{doi}
\usepackage{tikz}
\usepackage{lineno}
\usepackage[caption=false]{subfig}
\usepackage{breqn}
\usepackage[export]{adjustbox}
\usepackage{float}
\usepackage{placeins}

\usepackage{color}

\newcommand{\ii}{\text{i}}

\usepackage{braket}
\usepackage{stackengine}
\newcommand\xrowht[2][0]{\addstackgap[.5\dimexpr#2\relax]{\vphantom{#1}}}

\begin{document}
\begin{center}
{\Large \textbf{Universal correlations in the Abelian sandpile model}}
\end{center}

% TODO: write the author list here. Use initials + surname format.
% Separate subsequent authors by a comma, omit comma at the end of the list.
% Mark the corresponding author with a superscript *.
\begin{center}
Qiyu Liu\textsuperscript{1},
Jan-Niklas Herre\textsuperscript{2},
Prajit Baruah\textsuperscript{3},
Sebastian Dreizler\textsuperscript{1},\\
Christoph Karrasch\textsuperscript{1},
Wioletta Ruszel\textsuperscript{4} and
Dirk Schuricht\textsuperscript{3}
\end{center}

\begin{center}
{\bf 1} Technische Universit\"at Braunschweig, Institut f\"ur Mathematische Physik, Mendelssohnstrasse 3, 38106 Braunschweig, Germany\\
{\bf 2} Institute for Theory of Statistical Physics, RWTH Aachen University, Aachen, Germany\\
{\bf 3} Institute for Theoretical Physics, Utrecht University, Princetonplein 5, 3584 CE Utrecht, The Netherlands\\
{\bf 4} Mathematical Institute, Utrecht University, Budapestlaan 6, 3584 CD Utrecht, The Netherlands
\\
qiyu.liu@tu-braunschweig.de, d.schuricht@uu.nl
\end{center}

\begin{center}
\today
\end{center}

% For convenience during refereeing: line numbers
%\linenumbers

\section*{Abstract}
% abstract should be at most 8 lines.
{\bf We numerically study the bulk correlation functions in the two-dimensional Abelian sandpile model, aiming both to compare with the predictions of logarithmic conformal field theory and to extend the analysis to lattices where analytical methods are difficult to apply. Wilson's algorithm efficiently generates large-scale uniform spanning trees in parallel, which can be mapped to independent recurrent configurations via the Majumdar--Dhar burning bijection, eliminating sample autocorrelations and yielding fast convergence. On the square (single-sublattice) and honeycomb (two-sublattice) lattices our results agree well with the analytical predictions. For the kagome lattice we provide the first systematic numerical study of the bulk correlation functions, and cross-check the bulk height-1 probability against a closed-form analytical expression that we also derive here via the lattice Green function.}

\vspace{10pt}
\noindent\rule{\textwidth}{1pt}
\tableofcontents\thispagestyle{fancy}
\noindent\rule{\textwidth}{1pt}
\vspace{10pt}

%%%%%%%%%%%%%%%%%%%%%%%%%%%%
\section{Introduction}
\label{sec:intro}
%%%%%%%%%%%%%%%%%%%%%%%%%%%%

In statistical mechanics, criticality is characterised by a diverging correlation length $\xi \to \infty$ in the thermodynamic limit, leaving no preferred length scale and rendering the system scale invariant. This scale invariance is reflected in the power-law decay of correlation functions, $C(r) \sim r^{2-d-\eta}$, with $d$ denoting the system dimension. For equilibrium systems, criticality generically requires the fine-tuning of external parameters to special values. In contrast, scale-free behaviour is observed in many natural systems, from earthquake and forest fires statistics to solar flares. The contradiction between the requirement of fine-tuning in equilibrium theory and the widespread occurrence of scale-free behaviour in nature necessitated an alternative explanation beyond the equilibrium framework. Bak, Tang and Wiesenfeld~\cite{Bak-87,Bak-88} proposed the concept of \emph{self-organised criticality} (SOC), illustrating it with a cellular-automaton sandpile model in which open systems evolve spontaneously to a critical stationary state without any fine-tuning of external parameters. Dhar~\cite{Dhar90,Dhar06} later formalised and generalised this construction as the \emph{Abelian sandpile model} (ASM), establishing the Abelian property of the toppling operators that underlies its rich algebraic structure.

Many two-dimensional critical systems have, or are expected to have under suitable assumptions, conformally invariant scaling limits~\cite{BPZ84}, allowing the correlation functions to be computed within the framework of conformal field theory (CFT)~\cite{DiFrancescoMathieuSenechal97,Henkel99}. The scaling limit of the ASM is expected to be a \emph{logarithmic} CFT (LCFT)~\cite{GRR13} with central charge $c=-2$~\cite{Ruelle21}. Its logarithmic character is rooted in the non-local nature of the height variables when re-expressed in terms of the underlying spanning trees: the fluctuation of the minimally stable height, $h_1 - \langle h_1\rangle$, is identified with a Virasoro primary field of scaling dimension $\Delta=2$, while the higher-height fluctuation $h_2 - \langle h_2\rangle$ acts as its logarithmic partner, producing additional $\log r$ and $\log^2 r$ contributions to the corresponding two-point correlations~\cite{Ruelle21}. On the square lattice, exact asymptotic expressions for the corresponding two-point correlations have been obtained from graph-theoretic calculations of the underlying spanning-tree probabilities~\cite{MajumdarDhar92,PGPR08,PGPR10,PonceletRuelle17}, in agreement with the LCFT predictions~\cite{JPR06,PonceletRuelle17}.

These analytical results, however, rely heavily on the specific lattice structure: the derivations exploit the translational symmetry and known Green function of the square lattice. For more complex geometries---non-Bravais lattices, lattices with different coordination numbers, or irregular domains---the analytical calculations become significantly harder and the scaling-field coefficients are in general unknown. A robust numerical approach is therefore essential to provide insight into universality and lattice-dependent coefficients in regimes where analytical methods are difficult to apply.

The main numerical challenge lies in sampling. The standard sandpile dynamics, which involves random grain addition followed by toppling relaxation, generates strongly correlated successive configurations, requiring both a lengthy burn-in period to reach the stationary measure and the discarding of many intermediate configurations between successive samples to suppress autocorrelations. Previous studies used $\sim 10^9-10^{10}$ Markov-chain samples to obtain reliable fits for one-point functions in the presence of boundaries~\cite{PR05} as well as two-point functions in the bulk~\cite{JPR06}. Both were found to be in good agreement with the corresponding LCFT setups but considered only the square lattice. In this work, we bypass the Markov chain entirely by using Wilson's algorithm~\cite{Wilson96} to sample uniform spanning trees, combined with the Majumdar--Dhar burning bijection to produce recurrent sandpile configurations. Each sample is drawn independently, eliminating autocorrelations altogether. As a result, far fewer samples are needed to achieve comparable or better statistical precision: on a $8000 \times 8000$ square lattice, $10^6$ independent samples already suffice to obtain well-converged results. The absence of self-correlations yields cleaner data and faster convergence of estimators, allowing us to consider correlation functions in the ASM on the square, honeycomb and kagome lattices.

The remainder of this paper is organised as follows. Section~\ref{sec:model} defines the ASM, its recurrent configurations, and the burning bijection that maps them to spanning trees. Section~\ref{sec:obs} introduces the bulk observables of interest, summarises the analytical LCFT framework underpinning their predicted asymptotic forms, and describes our numerical implementation based on Wilson's algorithm and FFT-based correlation evaluation. Section~\ref{sec:universal} verifies the universal scaling exponent $2\Delta = 4$ on the three lattices studied here. Section~\ref{sec:results} then presents the lattice-by-lattice results: the square (Section~\ref{sec:square}), honeycomb (Section~\ref{sec:honeycomb}), and kagome (Section~\ref{sec:kagome}) lattices are treated in turn, in order of decreasing availability of closed-form analytical predictions, with Section~\ref{sec:CFTcheck} assessing the results in the LCFT framework. Section~\ref{sec:discussion} summarises the findings and discusses directions for future work.

%%%%%%%%%%%%%%%%%%%%%%%%%%%%
\section{The Abelian sandpile model}
\label{sec:model}
%%%%%%%%%%%%%%%%%%%%%%%%%%%%
\subsection{Definition}\label{sec:modeldefinition}
%%%%%%%%%%%%%%%%%%%%%%%%%%%%

\begin{figure}[t]
\centering
\includegraphics[width=0.9\textwidth]{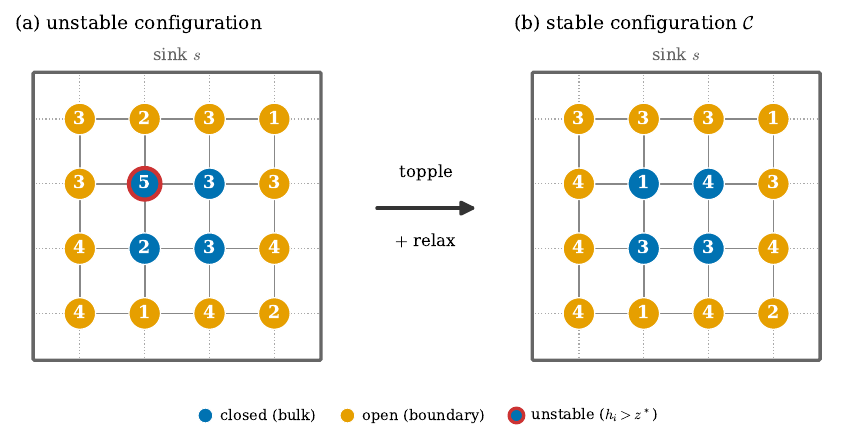}
\caption{Toppling dynamics of the ASM on a $4\times4$ square lattice with wired boundary conditions, ie, all boundary sites being connected to an additional vertex which we call the sink. Closed (bulk) sites are shown in blue, open (boundary) sites in orange, and the surrounding box represents the sink; dotted lines indicate edges connecting boundary sites to the sink. (a)~A stable configuration to which one grain has been added at an interior bulk site, which is thereby becomes unstable ($h_i = 5 > z_i^* = 4$, red ring). (b)~After relaxation the system reaches the stable configuration~$\mathcal{C}$: the unstable site has toppled (losing $z_i^*=4$ grains, dropping to $h=1$) and each of its four neighbours has gained one grain. This particular configuration~$\mathcal{C}$ is in fact \emph{recurrent}; its spanning-tree representation is analysed in Figure~\ref{fig:burning}.}
\label{fig:model_setup}
\end{figure}

Here we briefly review the definition of the ASM. Following the same notations introduced in Ruelle~\cite{Ruelle21}, we define the model on a finite connected graph~$\Gamma ({V}, {E})$, where ${V}$ and ${E}$ stand for vertices and edges on a grid with certain geometry. It is immediately clear that we can add dissipation to the system by introducing a new vertex $s$ and edges $o$ that connected a non-empty vertex subset $D \subset V$ to $s$ (marked as sink and dashed lines in Figure~\ref{fig:model_setup}). Then the full connected graph is defined as $\Gamma^{*} ({V}^{*}, {E}^{*})$, where ${V}^{*} = V \cup s$ and ${E}^{*} = E \cup o$. We can denote the coordination number of site~$i$ by~$z_i$ (resp.~$z_i^*$) in~$\Gamma$ (resp.~$\Gamma^*$), so that $z_i^* = z_i$ if~$i$ is closed and $z_i^* > z_i$ if~$i$ is open. To each site $i \in \Gamma$ we assign an integer-valued height variable $h_i \ge 1$. A height configuration $\mathcal{C} = \{h_i\}_{i\in V}$ is called \emph{stable} if
$h_i \le z_i^*$ for every site~$i$; otherwise site~$i$ is called \emph{unstable}.

The dynamics of the sandpile model is a Markov chain on stable configurations defined as follows. Starting from a stable configuration~$\mathcal{C}_t$ at time~$t$, the transition to $\mathcal{C}_{t+1}$ proceeds in two steps:
\begin{enumerate}
  \item \textbf{Addition.} A grain is added to a uniformly chosen random site~$i$, producing a new configuration $\mathcal{C}^{\mathrm{new}}$ with $h_i^{\mathrm{new}} = h_i + 1$ and all other heights unchanged. If $\mathcal{C}^{\mathrm{new}}$ is already stable, we set $\mathcal{C}_{t+1} = \mathcal{C}^{\mathrm{new}}$ and proceed to the next time step.
  \item \textbf{Relaxation.} If site~$i$ is unstable ($h_i^{\mathrm{new}} > z_i^*$), it \emph{topples}: each of its $z_i$ neighbours in~$\Gamma$ receives one grain, and the remaining $z_i^* - z_i$ grains are sent to the sink and removed from the system. This toppling may render one or   more neighbours unstable; each such site topples in turn by the same rule. The process repeats until all sites are stable, yielding the new configuration~$\mathcal{C}_{t+1}$.
\end{enumerate}
Dhar~\cite{Dhar90} proved that the final stable configuration is independent of the order in which unstable sites are toppled (the \emph{Abelian property}), so that~$\mathcal{C}_{t+1}$ is uniquely determined by~$\mathcal{C}_t$ and the addition site.

The toppling rule can be expressed compactly using the symmetric \emph{toppling matrix}~$\Delta$ on~$\Gamma$, defined by
\begin{equation}
  \Delta_{ij} =
  \begin{cases}
    z_i^* & \text{if } i=j, \\
    -1 & \text{if there exists an edge } \{i,j\}
          \text{ between } i \text{ and } j, \\
    0 & \text{otherwise.}
  \end{cases}
  \label{eq:toppling_matrix}
\end{equation}
When site~$i$ topples, the heights are updated as $h_j \mapsto h_j - \Delta_{ij}$ for all $j \in V$.

%%%%%%%%%%%%%%%%%%%%%%%%%%%%
\subsection{Recurrent configurations and spanning trees}
%%%%%%%%%%%%%%%%%%%%%%%%%%%%

The two-step dynamics of Section~\ref{sec:modeldefinition} can be encoded in a single \emph{operator}~$a_i$, which maps a stable configuration~$\mathcal{C}$ to the stable configuration $a_i(\mathcal{C})$ obtained by adding one grain at site~$i$ and relaxing to stability. The Abelian property proven by Dhar~\cite{Dhar90} states that $a_i \circ a_j = a_j \circ a_i$ for all sites~$i,j$, so the final result of a sequence of additions depends only on how many grains are added at each site, not on the order. An individual step in the Markov chain is thus $\mathcal{C}_{t+1} = a_i(\mathcal{C}_t)$ with~$i$ chosen uniformly at random, with the full Markov chain being constructed by repetition.

An important structural notion is that of \emph{forbidden subconfigurations} (FSCs)~\cite{Dhar90,MajumdarDhar92}. Let~$\mathcal{C}$ be a stable configuration and let~$F \subset \Gamma$ be a connected subgraph. The restriction of~$\mathcal{C}$ to~$F$ is a FSC if every vertex $i \in F$ satisfies $h_i \le n_i^F$, where $n_i^F$ is the number of neighbours of~$i$ that also belong to~$F$. No FSC can appear in the repeated image of the operators~$a_i$; consequently, a configuration is recurrent if and only if it contains no FSC. The simplest example is a pair of neighbouring vertices both with $h = 1$: each has one neighbour in~$F$, and $h = 1 \le 1 = n_i^F$. The criterion also implies that the maximal configuration $h_i = z_i^*$ for all~$i$ is always recurrent, since $h_i = z_i^* > n_i^F$ at the boundary for any proper subgraph~$F$. 

Stable configurations containing at least one FSC are \emph{transient}: once the dynamics leaves them, it can never return. Those free of any FSC are \emph{recurrent}: for every such configuration~$\mathcal{C}$ there exists a sequence of operators~$a_i$ that returns the system to~$\mathcal{C}$. Dhar~\cite{Dhar90,Dhar06} showed that the Markov chain possesses a unique invariant measure~$P_\Gamma$ whose support is precisely the set of recurrent configurations, and that $P_\Gamma$ is \emph{uniform} on this support.

The number of recurrent configurations admits a remarkably compact expression: Dhar~\cite{Dhar90} proved that $|\mathrm{Rec}| = \det(\Delta)$, where~$\Delta$ is the toppling matrix~\eqref{eq:toppling_matrix}. By Kirchhoff's matrix-tree theorem, $\det(\Delta)$ also equals the number of spanning trees of~$\Gamma^*$ rooted at the sink~$s$. This numerical coincidence suggests the existence of an explicit bijection between the two sets.

\begin{figure}[t]
\centering
\includegraphics[width=0.95\textwidth]{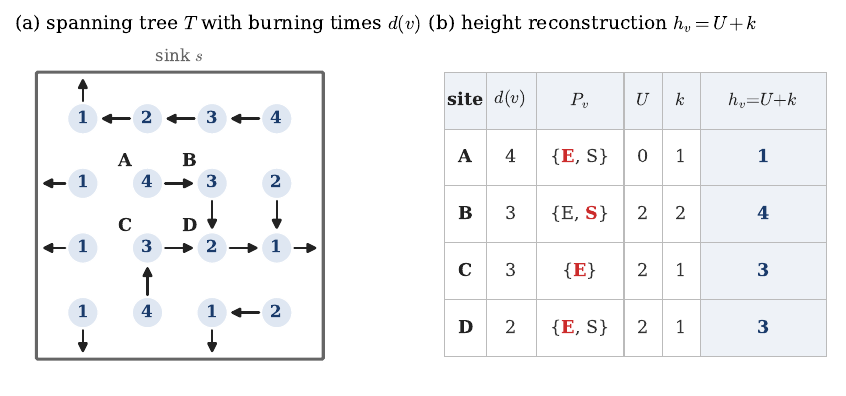}
\caption{The Majumdar--Dhar burning bijection for the recurrent configuration~$\mathcal{C}$ of Figure~\ref{fig:model_setup}(b). (a)~The sink-rooted spanning tree~$T=\sigma(\mathcal{C})$: arrows point from each site to its tree parent (towards the sink), and each site is labelled by its burning time~$d(v)$, equal to its depth in~$T$. Only the tree edges are drawn (as arrows); the four closed (bulk) sites are marked $A$--$D$. (b)~Reconstruction of the height at each bulk site. Reading the neighbours' burning times from panel~(a), $U$ counts the neighbours that burn at time~$\ge d(v)$. The column~$P_v$ lists the previous-layer neighbours (those in burning layer $d(v)-1$) as a subset of $\{E,N,W,S\}$ written in this fixed canonical order (one can analogously define orderings for the other lattices); the tree parent is shown in red, and $k$ is its rank within $P_v$. The height then follows as $h_v=U+k$. For example, site~$B$ ($d(v)=3$) has $U=2$ and $P_v=\{E,S\}$, whose second (red) element is the parent~$S$, giving $k=2$ and $h_B=4$. The reconstructed heights $h_v$ coincide with the values carried by the same sites in~$\mathcal{C}$ [Figure~\ref{fig:model_setup}(b)], confirming the bijection.}
\label{fig:burning}
\end{figure}
Such a bijection was constructed by Majumdar and Dhar~\cite{MajumdarDhar92} via a \emph{burning algorithm}. Given a recurrent configuration~$\mathcal{C}$ [Figure~\ref{fig:model_setup}(b)], all sites are initially declared \emph{unburned} while the sink burns at $t=0$. At each subsequent step~$t=1,2,\dots$, a site~$v$ ignites once the number of its still-unburned neighbours is at most $h_v-1$. At $t=1$ only the sink has burned, so this count equals the lattice coordination~$z_v$ defined in Section~\ref{sec:modeldefinition} and the condition reduces to
\begin{equation}
  h_v - 1 \;\ge\; z_v.
  \label{eq:burning_test}
\end{equation}
On the square lattice this gives the critical heights $h_v \ge 3$ at a corner ($z_v=2$) and $h_v \ge 4$ at an edge ($z_v=3$); a bulk site ($z_v = z_v^* = 4$) would need $h_v \ge 5$, which is impossible, so the fire necessarily starts at the boundary and propagates inward as more neighbours become burned. A configuration is recurrent if and only if all sites eventually burn, and the burning time~$d(v)=t$ at which~$v$ ignites equals its depth in the resulting spanning tree~$T = \sigma(\mathcal{C})$.

When~$v$ ignites it is connected by a tree edge to a unique parent in the preceding burning layer. To specify this parent we introduce two auxiliary quantities. Let
\begin{equation}
  U \;=\; \#\{\,\text{neighbours of $v$ with burning time }\ge d(v)\,\}
  \label{eq:U_def}
\end{equation}
denote the number of neighbours of~$v$ still unburned at the moment of ignition, and let~$P_v$ denote the set of neighbours of~$v$ with burning time exactly $d(v)-1$ (the \emph{previous layer}), enumerated in a fixed canonical order, here $E,N,W,S$. The tree parent of~$v$ is then the $k$-th element of~$P_v$, where the rank
\begin{equation}
  k \;=\; h_v - U \;\in\; \{1,\dots,|P_v|\}
  \label{eq:parent_rank}
\end{equation}
is fixed by $h_v$ and $U$. If $|P_v|=1$ the choice is trivially unique. Applying this procedure to the recurrent configuration~$\mathcal{C}$ of Figure~\ref{fig:model_setup}(b) yields the spanning tree~$T$ shown in Figure~\ref{fig:burning}(a), each site labelled by its burning time~$d(v)$.

This map is invertible, which is in fact the direction relevant for our numerical sampling: Wilson's algorithm (Section~\ref{sec:algorithm}) returns a uniform spanning tree~$T$ but no heights. Given~$T$, the burning times are obtained directly as the tree depths,
\begin{equation}
  d(v) \;=\; \text{(graph distance from~$v$ to the sink~$s$ along $T$)},
  \label{eq:depth}
\end{equation}
by a single sweep of the burning process starting from the sink. From these depths, $U$ is recovered as in~\eqref{eq:U_def} (the number of neighbours of~$v$ with $d \ge d(v)$), $P_v$ as the subset of neighbours with $d = d(v)-1$ in canonical order $E,N,W,S$, and $k$ as the rank within~$P_v$ of the actual tree parent of~$v$ read off from~$T$. The height at every site then follows from reading~\eqref{eq:parent_rank} as
\begin{equation}
  h_v \;=\; U + k.
  \label{eq:height_reconstruction}
\end{equation}
Figure~\ref{fig:burning}(b) tabulates $d(v)$, $P_v$, $U$, $k$ and the reconstructed~$h_v$ for the four bulk sites $A$--$D$ of the tree in Figure~\ref{fig:burning}(a); the heights agree with those of~$\mathcal{C}$ [Figure~\ref{fig:model_setup}(b)], establishing the bijection $T \leftrightarrow \mathcal{C}$.

The correspondence has a natural interpretation: both spanning trees and recurrent configurations encode global connectivity to the sink without frozen regions. A spanning tree provides every vertex with a unique path to the sink, while a recurrent configuration contains no FSC---no cluster of sites is so depleted that it can never be activated by avalanches from the boundary. The burning process bridges these viewpoints: the tree dictates the order in which fire propagates from the sink, and the height at each site records how many neighbours remain unburned together with the identity of the tree parent. This bijection underpins the sampling strategy described in Section~\ref{sec:algorithm}.

%%%%%%%%%%%%%%%%%%%%%%%%%%%%
\subsection{Lattices studied}
%%%%%%%%%%%%%%%%%%%%%%%%%%%%

\begin{figure}[t]
\centering
\includegraphics[width=0.95\textwidth]{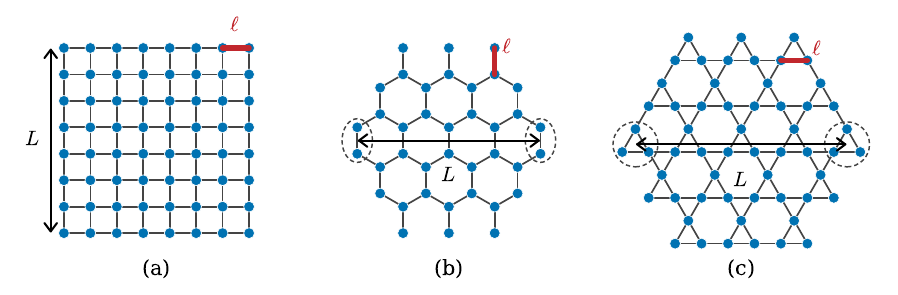}
\caption{The three planar lattices studied in this work.  (a)~Square lattice ($z=4$).  (b)~Honeycomb lattice ($z=3$).  (c)~Kagome lattice ($z=4$); the honeycomb and kagome domains are hexagonal, reflecting the natural boundary of the underlying triangular Bravais lattice.  In each panel $L$ is the linear system size (the domain diameter for the hexagonal cases), measured in nearest-neighbour (bond-length $\ell$) units.  A single nearest-neighbour bond of length $\ell$ is highlighted (red) in each panel; we set $\ell=1$ throughout, so all distances and system sizes are quoted in these units.  Writing $a$ for the lattice constant ($a=\ell$ square, $a=\sqrt{3}\,\ell$ honeycomb, $a=2\,\ell$ kagome), the simulated linear sizes are $L=8000\,\ell$ (square), $L=6000\sqrt{3}\,\ell$ (honeycomb) and $L=6000\,(2\ell)=12000\,\ell$ (kagome).}
\label{fig:lattices}
\end{figure}

Figure~\ref{fig:lattices} shows the three planar lattices studied in this work. Each lattice is characterised by its coordination number~$z$ (the number of nearest neighbours per bulk site) and the number of sites per unit cell:
\begin{itemize}
\item The \textbf{square lattice} ($z=4$): a Bravais lattice with one site per unit cell and heights $h\in\{1,2,3,4\}$. We use square domains of side length~$L$ with wired boundary conditions (all boundary sites connected to the sink).
\item The \textbf{honeycomb lattice} ($z=3$): two sites per unit cell on a triangular Bravais lattice, with heights $h\in\{1,2,3\}$. We use hexagonal domains of radius~$R$ with wired boundaries.
\item The \textbf{kagome lattice} ($z=4$): a lattice of corner-sharing triangles with three sublattices (A, B, C) per unit cell and heights $h\in\{1,2,3,4\}$. We use hexagonal domains of radius~$R$ with wired boundaries.
\end{itemize}
These three lattices were chosen to span a natural progression in both analytical control and structural complexity. The square lattice serves as a benchmark case: many of the relevant coefficients are known exactly (see Section~\ref{sec:square_predictions}), making it an ideal testbed for validating our numerical approach. The honeycomb lattice is less well explored, but still allows comparison with several quantities of interest; at the same time, its two-sublattice structure already introduces a non-Bravais complication absent on the square lattice. Finally, the kagome lattice is much less understood: apart from the exact average density, few bulk observables are known analytically. In the case of the kagome lattice, our numerical results therefore play a predictive role, providing estimates for one-point height probabilities and connected two-point correlation functions that may help clarify the physics of this lattice and guide future analytical work.

%%%%%%%%%%%%%%%%%%%%%%%%%%%%
\section{Observables: analytical predictions and numerical implementation}
\label{sec:obs}
%%%%%%%%%%%%%%%%%%%%%%%%%%%%

The main focus of this and the following sections is the computation of bulk observables on the static support of the recurrent configurations. After defining the observables of interest, we briefly summarise the analytical LCFT framework that governs their asymptotic behaviour, and then describe our numerical implementation. The lattice-specific predictions, including the explicit closed-form values of the LCFT coefficients on the square lattice, are presented together with the corresponding numerical analysis in Section~\ref{sec:square}.

%%%%%%%%%%%%%%%%%%%%%%%%%%%%
\subsection{Observables of interest}
\label{sec:observables}
%%%%%%%%%%%%%%%%%%%%%%%%%%%%

Our analysis focuses on two classes of bulk observables: one-point height probabilities and connected two-point correlation functions. The one-site probability is defined by
\begin{equation}
  P_a = \lim_{|V|\to\infty} \mathbb{P}(h_i = a),
  \label{eq:siteprobabilities}
\end{equation}
where the limit is taken for bulk sites far from the boundary.

To study correlations, we introduce the subtracted height variables $h_a(\mathbf{x}) = \delta_{h(\mathbf{x}),a} - P_a$, which are normalised such that $\langle h_a(\mathbf{x})\rangle=0$ in the bulk. The connected two-point function is now defined as 
\begin{equation}
  C_{a,b}(\mathbf{r}) = \Big\langle h_{a}(\mathbf{x})\, h_{b}(\mathbf{x}+\mathbf{r}) \Big\rangle
  = \mathbb{P}\Bigl(h(\mathbf{x})=a,\, h(\mathbf{x}+\mathbf{r})=b\Bigr) - P_{a} P_{b},
  \label{eq:corr_def}
\end{equation}
where the expectation value $\langle\ldots\rangle$ is taken over all recurrent configurations. The coordinates $\mathbf{x}$ and $\mathbf{x}+\mathbf{r}$ are assumed to be sufficiently far from the boundary such that the correlations only depend on their difference $\mathbf{r}$. 
In practice, we work with the radial correlation $C_{a,b}(r)$ obtained by averaging over all pairs at Euclidean distance $r=|\mathbf{r}|$. See Section~\ref{sec:algorithm} below for further details.

Note that the connected correlation functions $C_{a,b}(r)$ require knowledge of the one-point probabilities $P_a$ entering through the  subtracted height variables $h_a$. When evaluating the analytical predictions discussed below, the exact values of $P_a$ are used directly. In our numerical analysis, however, we systematically use the empirical estimates of $P_a$ obtained from the same set of samples used to compute $C_{a,b}(r)$. This ensures that the subtracted height variables have a vanishing sample mean by construction and that the resulting correlation functions are self-consistent, without importing analytical input into the numerical pipeline.

%%%%%%%%%%%%%%%%%%%%%%%%%%%%
\subsection{Analytical predictions: LCFT framework}
\label{sec:CFT}
%%%%%%%%%%%%%%%%%%%%%%%%%%%%

As was established by Ruelle and collaborators (see Reference~\cite{Ruelle21} for a review), the scaling limit of the ASM is described by a LCFT with central charge $c=-2$. The negative central charge indicates non-unitarity of the theory, here being a consequence of the presence of dissipation (via the sink) in the ASM. The defining feature of an LCFT is the appearance of indecomposable Jordan blocks in the action of the dilation generator $L_0$, resulting in characteristic logarithmic corrections in correlation functions. 

For the description of the scaling limit of the correlation functions \eqref{eq:corr_def} two Jordan blocks are relevant: The first is formed by the identity field $\mathbb{I}$ and the dissipation field $\omega$, both with scaling dimensions $\Delta=0$. The second block is formed by the field $\phi(z,\bar{z})$ and its logarithmic partner $\psi(z,\bar{z})$ with scaling dimensions $\Delta=2$, where we used complex coordinates $z$ and $\bar{z}$. The relevant correlation functions are given by 
\begin{eqnarray}
C_{\phi\phi\omega}(\mathbf{r})&=&\Big\langle \phi(z_1,\bar{z}_1)\,\phi(z_2,\bar{z}_2)\,\omega(\infty)\Big\rangle_{\mathrm{LCFT}}\simeq\frac{A}{r^4},\label{eq:LCFTcorrelator1}\\
C_{\phi\psi\omega}(\mathbf{r})&=&\Big\langle \phi(z_1,\bar{z}_1)\,\psi(z_2,\bar{z}_2)\,\omega(\infty)\Big\rangle_{\mathrm{LCFT}}\simeq\frac{A\log r+B}{r^4},\label{eq:LCFTcorrelator2}\\
C_{\psi\psi\omega}(\mathbf{r})&=&\Big\langle \psi(z_1,\bar{z}_1)\,\psi(z_2,\bar{z}_2)\,\omega(\infty)\Big\rangle_{\mathrm{LCFT}}\simeq\frac{A\log^2r+2B\log r+D}{r^4},
\label{eq:LCFTcorrelator3}
\end{eqnarray}
where the expectation value is taken in the LCFT and $|z_2-z_1|=|\mathbf{r}|=r$. The insertion of the dissipation field $\omega(\infty)$ at infinity represents the presence of the sink. In the last step we have stated the leading behaviours at large distances $r$, with the constants $A$, $B$ and $D$ being set by the normalisation. Throughout our manuscript we fix $A=-P_1^2/2$ with the one-site probability \eqref{eq:siteprobabilities} for all considered lattices. We note that the correlation functions contain characteristic logarithmic contributions. The pair $\{\phi,\psi\}$ forms a staggered module of the Virasoro algebra~\cite{KytolaRidout09}, which cannot be represented in terms of usual fermionic fields~\cite{PonceletRuelle17,Ruelle21}. 

The height variables are now obtained as linear combinations~\cite{PR05,JPR06,Ruelle21} 
\begin{equation}
  h_a(z,\bar{z}) = \alpha_a\,\psi(z,\bar{z})  + \beta_a\,\phi(z,\bar{z}),
  \label{eq:field_decomp}
\end{equation}
where $\alpha_1=\beta_2=0$, with the other parameters $\alpha_a$ and $\beta_a$ being lattice dependent and thus non-universal (see Section~\ref{sec:results} below). In this setting the scaling limit of the correlation function \eqref{eq:corr_def} can be expressed as combinations of the correlators \eqref{eq:LCFTcorrelator1}--\eqref{eq:LCFTcorrelator3}, yielding the universal asymptotic form of the connected two-point function on any sufficiently isotropic, planar lattice,
\begin{equation}
\begin{split}
  r^4\,C_{a,b}(r) \simeq &\alpha_a\alpha_b A\,\log^2 r + \bigl[2\alpha_a\alpha_b B+(\alpha_a\beta_b+\alpha_b\beta_a)A\bigr]\,\log r\\ 
  &\quad + \alpha_a\alpha_b D+(\alpha_a\beta_b+\alpha_b\beta_a)B+\beta_a\beta_b A.
  \end{split}
  \label{eq:Cab_theory}
\end{equation}
The leading behaviours thus read 
\begin{equation}
C_{1,1}(r)\simeq\frac{\beta_1^2A}{r^4},\quad C_{a,1}(r)\simeq \frac{\alpha_a\beta_1A}{r^4}\log r,\quad C_{a,b}(r)\simeq\frac{\alpha_a\alpha_bA}{r^4}\log^2 r\quad\text{for}\; a,b\ge 2.
\label{eq:leadingCF}
\end{equation}
Below we use the notation $\mathcal{A}_{ab}$ for the corresponding amplitudes, ie, $\mathcal{A}_{11}=\beta_1^2A$, $\mathcal{A}_{a1}=\alpha_a\beta_1A$ and $\mathcal{A}_{ab}=\alpha_a\alpha_bA$. For the square lattice all of the amplitudes above can be obtained in closed form by exploiting the underlying spanning-tree representation~\cite{PR05,JPR06,PonceletRuelle17}. In particular, in the present normalisation, ie, with $A=-P_1^2/2$, the expansion parameters $\alpha_a$, $\beta_a$ can be obtained analytically with the result $\alpha_2=\beta_1=1$ and the others stated in Section~\ref{sec:square_predictions} below. On the honeycomb and the kagome lattice the coefficients $\alpha_a$, $\beta_a$ are not known analytically (apart from limited information on the height-1 sector); the universal form~\eqref{eq:Cab_theory} nevertheless applies, and the resulting lattice-dependent amplitudes are among the quantities extracted from our numerical data in Sections~\ref{sec:honeycomb} and~\ref{sec:kagome}.

%%%%%%%%%%%%%%%%%%%%%%%%%%%%
\subsection{Numerical implementation}
\label{sec:algorithm}
%%%%%%%%%%%%%%%%%%%%%%%%%%%%

\paragraph{Sampling.}
Our sampling strategy bypasses the correlated Markov-chain dynamics entirely. Instead, we exploit the exact bijection between recurrent sandpile configurations and sink-rooted spanning trees. The algorithm proceeds in three steps:
\begin{enumerate}
  \item \textbf{Wilson's algorithm}: Sample a uniform spanning tree (UST) by performing loop-erased random walks (LERW) from each vertex to the growing tree~\cite{Wilson96}. This produces an exactly uniform sample from the set of all sink-rooted spanning trees.
  \item \textbf{Burning bijection}: Convert the spanning tree to a recurrent sandpile configuration using the Majumdar--Dhar burning test, which assigns heights based on the tree structure.
  \item \textbf{Measurement}: Measurement of observables on the generated configuration. 
  \item \textbf{Repeat}: Each call to Wilson's algorithm produces an \emph{independent} recurrent configuration, avoiding autocorrelation issues entirely.
\end{enumerate}
This approach is particularly advantageous for large lattices. For a square lattice of size $L$, Wilson's algorithm generates one sample in $O(L^2)$ expected time, while direct sandpile dynamics requires $O(L^2)$ steps per addition and relaxation cycle, with an autocorrelation time that grows with the system size. In our implementation, we generate $n_{\text{total}}\sim 10^6$ independent samples on a $L=8000\,\ell$ square lattice using 48 parallel CPU threads. Similar simulations are performed for the other lattices, with the precise details stated next to the shown results.

\paragraph{Correlation function evaluation.}
To efficiently compute the radially averaged correlation function~\eqref{eq:corr_def}, we use the convolution theorem. For each sample, we form the centered indicator field $J_a(\mathbf{x}) = \delta_{h(\mathbf{x}),a} - {P}_a$ on the interior of the lattice (excluding a margin of width $m\sim 200$ near the boundary to suppress finite-size effects), where ${P}_a$ is the empirical estimate of $P_a$ from the same sample set, as discussed in Section~\ref{sec:observables}. The two-point function for a given shift $\mathbf{d}$ is then
\begin{equation}
  {C}_{a,b}(\mathbf{d}) = \frac{1}{(N-|d_x|)(N-|d_y|)} \sum_{\mathbf{x}} J_{a}(\mathbf{x})\, J_{b}(\mathbf{x}+\mathbf{d}),
\end{equation}
where $N = L - 2m$ is the interior window size. This sum is computed via zero-padded FFT (padding to size $P \ge N + \rho$) to ensure linear (non-circular) convolution for all shifts up to $\rho$. The normalisation by $(N-|d_x|)(N-|d_y|)$ accounts for the exact number of non-wrapping pairs at each shift. The shift-resolved correlations are then binned by radial distance $r = |\mathbf{d}|$ and averaged over all samples to obtain $C_{a,b}(r)$.

More specifically, we use $\rho=30$ for simulations on the square lattice, so only radial correlations within $r\in[1,30]$ are retained. (On the hexagonal and kagome lattices $\rho$ is given by $30\sqrt{3}\approx 52$ and $60$, respectively.) The FFT step first produces the correlation as a function of the full displacement vector $\mathbf{d}=(d_x,d_y)$, denoted by ${C}_{a,b}(\mathbf{d})$. We then convert this displacement-resolved correlation into a radial correlation by grouping integer displacements into bins according to
\begin{equation}
    r=\mathrm{round}\left(\sqrt{d_x^2+d_y^2}\right).
\end{equation}
Thus, all displacements whose spatial lengths round to the same integer $r$ contribute to the same radial bin. For example, $(10,0)$, $(8,6)$, and symmetry-related displacements are all assigned to the $r=10$ bin. In practice, we precompute all integer displacements $(d_x,d_y)$ with $-\rho\le d_x,d_y\le\rho$ whose rounded distance lies in $1\le r\le \rho$, together with their corresponding radial bin. After the FFT routine, we got the total contribution to ${C}_{a,b}(\mathbf{d})$ for each displacement in each of the bins. Their contributions are accumulated in the appropriate radial bins and then normalised by the total number of valid non-wrapping site pairs. This yields the radially averaged correlation without explicitly enumerating individual site pairs.

\paragraph{Fitting strategy.}
The analysis of the numerical correlation functions proceeds in two stages, which we have organised into the structure of the paper as follows. (i)~The \emph{universal} power-law exponent $2\Delta = 4$ in Eq.~\eqref{eq:leadingCF} is first verified on all three lattices by fitting $-C_{1,1}(r) = A/r^{2\Delta}$ with $A$ and $\Delta$ free; this consolidated cross-lattice check is presented in Section~\ref{sec:universal}. (ii)~With the prefactor $r^4$ fixed, the \emph{lattice-dependent} amplitudes and sub-leading coefficients in Eq.~\eqref{eq:Cab_theory} are then extracted lattice by lattice and compared with the theoretical predictions where available; this is the content of Section~\ref{sec:results}.

%%%%%%%%%%%%%%%%%%%%%%%%%%%%
\section{Universal scaling exponent across all lattices}
\label{sec:universal}
%%%%%%%%%%%%%%%%%%%%%%%%%%%%

\begin{figure}[t]
\centering
\includegraphics[width=0.95\textwidth]{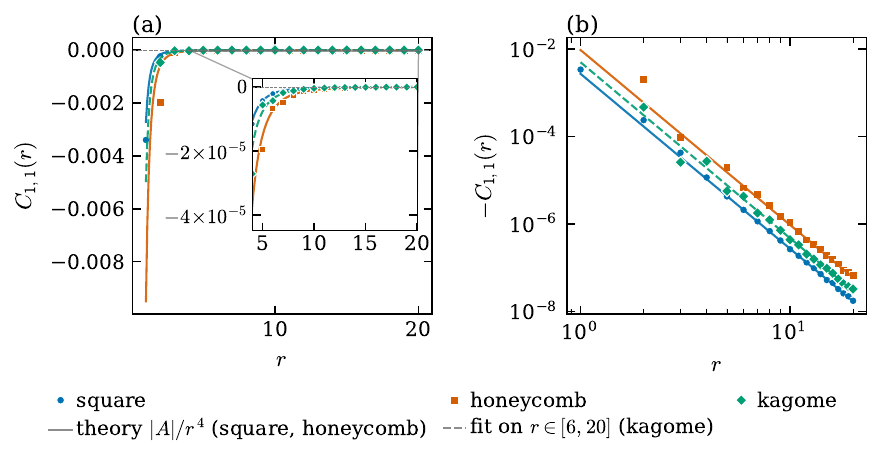}
\caption{Universal $C_{1,1}(r)\sim r^{-4}$ decay on the square, honeycomb and kagome lattices ($r=1-20$, nn/bond-length $\ell$ units).  Honeycomb/kagome curves are sublattice-averaged.  Solid lines: exact $A/r^4$ for square ($A^{\mathrm{sq}}=-P_1^2/2$) and honeycomb ($A^{\mathrm{hon}}=-9/(96\pi^2)$ in $\ell$-units); dashed: log-log power-law fit on $r\in[6,20]$ (kagome).  (a)~$C_{1,1}$ vs $r$.  (b)~$-C_{1,1}$ vs $r$ (log-log).}
\label{fig:universal_h1h1}
\end{figure}

A first universal prediction is that the connected two-point function of the minimally stable height obeys a pure power law $C_{1,1}(r)\sim r^{-2\Delta}$ with the lattice-independent exponent $2\Delta = 4$~\cite{MajumdarDhar91,Ruelle21}. Before turning to the lattice-dependent amplitudes in Section~\ref{sec:results}, we test this universality directly on the three lattices studied here by fitting
\begin{equation}
  -C_{1,1}(r) = \frac{A}{r^{2\Delta}}
\end{equation}
with both $A$ and $\Delta$ left free, and---crucially---on the \emph{same} fit window $r \in [r_\min,r_\max]=[6,20]$ for all three lattices. This common window strikes a balance between excluding the strongly lattice-dependent short-distance regime $r\lesssim 3$ and avoiding the noisier large-$r$ tail. For the honeycomb and kagome lattices the data are first averaged over the two and three sublattices, respectively, before the fit. Figure~\ref{fig:universal_h1h1} shows $C_{1,1}(r)$ on the three lattices both on a linear scale [panel~(a)] and on a log-log scale [panel~(b)], the latter including the per-lattice fits and a $r^{-4}$ guide line for reference. The numerical results are collected in Table~\ref{tab:universal_exponent}.

\begin{table}[t]
\centering
\begin{tabular}{lccc}
  \toprule
  lattice & fitted $2\Delta$ & deviation from 4 & mean inter-sublattice spread \\
  \midrule
  square    & $3.983 \pm 0.021$ & $-0.4\%$ & --- \\
  honeycomb & $4.009 \pm 0.058$ & $+0.2\%$ & $2.84 \times 10^{-9}$ \\
  kagome    & $4.024 \pm 0.057$ & $+0.6\%$ & $2.43 \times 10^{-9}$ \\
  \bottomrule
\end{tabular}
\caption{Fitted scaling exponent $2\Delta$ on a \emph{common} window $r\in[r_\min,r_\max]=[6,20]$ across all three lattices, obtained from $-C_{1,1}(r) = A/r^{2\Delta}$ by log-log linear regression. The reported uncertainty is the standard error of the slope from the regression. 
Square: $L=8000\,\ell$ ($n_{\mathrm{total}}\approx10^6$); honeycomb: $L=6000\sqrt{3}\,\ell$ ($n_{\mathrm{total}}\approx10^6$, averaged over 2 sublattices); kagome: $L=12000\,\ell$ ($n_{\mathrm{total}}\approx1.2\times10^6$, averaged over 3 sublattices).  The mean inter-sublattice spread is the average over the fit window of the standard deviation of $C_{1,1}(r)$ across sublattices.}
\label{tab:universal_exponent}
\end{table}

On all three lattices the fitted exponent agrees with the universal value $2\Delta = 4$ to within about one percent, and the residual deviations are statistically consistent with the slope-uncertainty of the same regression ($\sigma_{2\Delta}\sim 0.02-0.06$). For the honeycomb and kagome lattices the inter-sublattice spread of $C_{1,1}(r)$ stays at the $10^{-9}$ level across the fit window, confirming that the non-Bravais sublattice structure does not produce any visible splitting of the height-1 correlator. These observations validate the universal $~r^{-4}$ decay for the bulk correlations on all three lattices, and justify fixing the prefactor $r^{2\Delta} = r^4$ when extracting the lattice-dependent amplitudes in the next three sections.

%%%%%%%%%%%%%%%%%%%%%%%%%%%%
\section{Results for the three lattices}
\label{sec:results}
%%%%%%%%%%%%%%%%%%%%%%%%%%%%

Having confirmed the universal scaling exponent $2\Delta = 4$ in Section~\ref{sec:universal}, we now turn to the lattice-specific quantities. For each lattice we present, in turn, (i)~the one-point height probabilities $P_a$, compared with the analytical predictions where available, and (ii)~the connected two-point correlation functions $C_{a,b}(r)$, from which the lattice-dependent amplitudes $A$ and sub-leading constants of the LCFT decomposition~\eqref{eq:Cab_theory} are extracted. The square lattice (Section~\ref{sec:square}) serves as the most stringent benchmark, since closed-form predictions are available for all the coefficients we measure. On the honeycomb lattice (Section~\ref{sec:honeycomb}) only the leading height-1 amplitude $A^{\mathrm{hon}} = -1/(96\pi^2)$ is known analytically~\cite{Honeycomb10}, while on the kagome lattice (Section~\ref{sec:kagome}) we additionally provide a closed-form derivation of the bulk height-1 probability $P_1$ via the lattice Green function---to our knowledge the first such result reported for this geometry---and then proceed with the numerical analysis of the higher-height observables.

%%%%%%%%%%%%%%%%%%%%%%%%%%%%
\subsection{Square lattice}
\label{sec:square}
%%%%%%%%%%%%%%%%%%%%%%%%%%%%

We study the ASM on square domains of side length~$L$ with wired boundary conditions [see Figure~\ref{fig:lattices}(a) for the geometry and the definition of~$L$].

%%%%%%%%%%%%%%%%%%%%%%%%%%%%
\subsubsection{Height probabilities}
%%%%%%%%%%%%%%%%%%%%%%%%%%%%

Table~\ref{tab:square_heights} compares the measured one-point height probabilities with the exact analytical predictions (see~\cite{Ruelle21})
\begin{equation}
\begin{split}
&P_1=\frac{2}{\pi^2}-\frac{4}{\pi^3},\quad
P_2=\frac{1}{4}-\frac{1}{2\pi}-\frac{3}{\pi^2}+\frac{12}{\pi^3},\\[4pt]
&P_3=\frac{3}{8}+\frac{1}{\pi}-\frac{12}{\pi^3},\quad
P_4=\frac{3}{8}-\frac{1}{2\pi}+\frac{1}{\pi^2}+\frac{4}{\pi^3}.
\end{split}
\end{equation}
On a $L=8000\,\ell$ lattice with margin $m=201$ and $10^6$ samples, the agreement is at the $10^{-6}$ level.
\begin{table}[t]
\centering
\begin{tabular}{lccc}
  \toprule
  height & analytic prediction & simulation & rel.\ error \\
  \midrule
  $P_1$ & $0.073636230$ & $0.073636312$ & $1.125 \times 10^{-6}$ \\
  $P_2$ & $0.173899919$ & $0.173900288$ & $2.120 \times 10^{-6}$ \\
  $P_3$ & $0.306291473$ & $0.306292060$ & $1.916 \times 10^{-6}$ \\
  $P_4$ & $0.446172378$ & $0.446171340$ & $2.327 \times 10^{-6}$ \\
  \bottomrule
\end{tabular}
\caption{Square lattice height probabilities. Simulation: $L=8000\,\ell$, margin $m=201$, $n_{\mathrm{total}}\approx10^6$ samples.}
\label{tab:square_heights}
\end{table}

%%%%%%%%%%%%%%%%%%%%%%%%%%%%
\subsubsection{Two-point correlation functions}
%%%%%%%%%%%%%%%%%%%%%%%%%%%%

\paragraph{Theoretical predictions from the LCFT framework.}
\label{sec:square_predictions}
For the square lattice, the scaling-field coefficients $\alpha_a, \beta_a$ entering the LCFT decomposition~\eqref{eq:field_decomp} are known in closed form~\cite{JPR06}: 
\begin{align}
  &\alpha_1 = 0, \; \alpha_2 = 1, \;
  \alpha_3 = \frac{8-\pi}{2(\pi-2)} \approx 2.128, \;
  \alpha_4 = -\frac{\pi+4}{2(\pi-2)} \approx -3.128, \label{eq:alpha_square}\\[4pt]
  &\beta_1 = 1, \; \beta_2 = 0, \;
  \beta_3 = \frac{\pi^3 - 5\pi^2 + 12\pi - 48}{4(\pi-2)^2} \approx -5.495, \;
  \beta_4 = \frac{32 + 4\pi + \pi^2 - \pi^3}{4(\pi-2)^2} \approx 4.495, \label{eq:beta_square}
\end{align}
and they satisfy $\sum_a \alpha_a = \sum_a \beta_a = 0$, as required by the constraint $\sum_a h_a = 0$ on the height variables. Substituting these coefficients into~\eqref{eq:leadingCF} yields the leading amplitudes $\mathcal{A}_{ab}=\alpha_a\alpha_b A$ that we will compare to the fitted values below. For the cases involving height~1, the sub-leading shift constants are also known exactly~\cite{PonceletRuelle17}:
\begin{align}
  & r^4C_{a,1}(r)=\mathcal{A}_{a1}\bigl(\log r+c_{a1}\bigr)+o(1),\\[4pt]
  &c_{21} = \gamma + \frac{3}{2}\log 2 + \frac{16-5\pi}{2(\pi-2)} \approx 1.7448,
  \;
  c_{31} = \gamma + \frac{3}{2}\log 2 - \frac{40-2\pi-\pi^2}{2(8-\pi)} \approx -0.8373,
  \label{eq:c_square}
\end{align}
with $\gamma \approx 0.5772$ the Euler--Mascheroni constant. These predictions, together with the exact $P_1 = 0.073636$, fully determine the universal LCFT picture against which we benchmark our numerical analysis in the following.

\begin{figure}[t]
\centering
\includegraphics[width=0.95\textwidth]{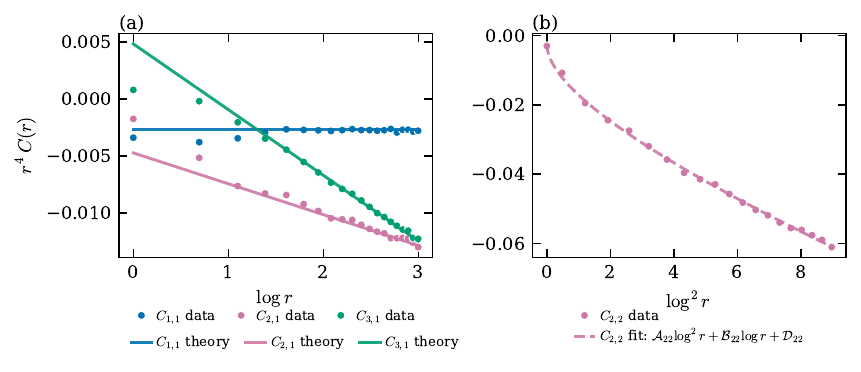}
\caption{Square lattice rescaled correlations ($L=8000\,\ell$; $\ell=a$).  (a)~$r^4C(r)$ vs $\log r$ for $C_{1,1}$, $C_{2,1}$, $C_{3,1}$ with exact theory lines \eqref{eq:Cab_theory} using the coefficients \eqref{eq:alpha_square} and~\eqref{eq:beta_square} for the leading long-distance behaviour ($n_{\text{total}}\approx10^6$ samples).  (b)~$r^4C_{2,2}(r)$ vs $\log^2 r$ with three-parameter fit $\mathcal{A}_{22}\log^2 r+\mathcal{B}_{22}\log r+\mathcal{D}_{22}$ ($n_{\text{total}}\approx10^6$, fit window $r\in[1,20]$).}
\label{fig:square_corr}
\end{figure}

\paragraph{Extraction of lattice-dependent amplitudes.}
\label{sec:square_amplitudes}
Having confirmed the universal exponent $2\Delta = 4$ in Section~\ref{sec:universal}, we now fix the power-law prefactor $r^4$ and extract the lattice-dependent amplitudes and sub-leading coefficients on the square lattice. Figure~\ref{fig:square_corr} presents the rescaled correlations $r^4\,C(r)$. Panel~(a) shows $r^4\,C(r)$ vs.\ $\log r$ for $C_{1,1}$, $C_{2,1}$ and $C_{3,1}$ ($n_{\text{total}}=10^6$), compared with the theoretical predictions from Section~\ref{sec:CFT}. Panel~(b) displays $r^4\,C_{2,2}(r)$ vs.\ $\log^2 r$ ($n_{\text{total}}=10^6$), fitted to Eq.~\eqref{eq:Cab_theory}.

Table~\ref{tab:square_fits} summarises the leading amplitudes $\mathcal{A}_{ab}$ extracted from the fits, compared side-by-side with the exact LCFT predictions obtained by substituting~\eqref{eq:alpha_square}-\eqref{eq:beta_square}. For the correlations $C_{a,1}$ we choose the fit window $r\in[6,20]$. For the three-parameter fits of $C_{2,2}$, we use the widest reliable intervals available: $r\in[1,20]$ for the square lattice and $r\in[2,20]$ for the honeycomb and kagome lattices (see below). Separating the $\log^2 r$, $\log r$ and constant contributions requires a broad logarithmic range of width $\log(r_\max/r_{\min})$. Although increasing $r_\min$ suppresses short-distance lattice corrections, it also makes the three fitted coefficients more strongly correlated and the fit more sensitive to noisy large-distance data, leading to stronger fluctuations as $r_\max$ is varied. The chosen windows provide the most stable compromise among those tested. All obtained amplitudes agree with theory to within about $12\%$. 

For $C_{2,2}$, a simpler 2-parameter fit (ignoring the sub-leading $\mathcal{B}_{22}\log r$ term) gives a dramatically biased estimate of $\mathcal{A}_{22}$ ($\sim 116\%$ deviation), demonstrating that the $O(\log r)$ correction is essential even for $r \lesssim 20$. The three-parameter fit agrees with the exact leading amplitude to within $11.3\%$, although its coefficient uncertainty remains appreciable because the leading $\log^2 r$ term is weak compared with the sub-leading correction.

\begin{table}[t]
\centering
\begin{tabular}{llccc}
  \toprule
  $r^4\times$ & fit form  & $\mathcal{A}_{ab}^{\text{fit}}$ & $\mathcal{A}_{ab}^{\text{theory}}$ & ratio \\
  \midrule
  $C_{1,1}$ & $\mathcal{A}_{11}$ & $(-2.752 \pm 0.021) \times 10^{-3}$ & $-2.711 \times 10^{-3}$ & $1.015$ \\[2pt]
  $C_{2,1}$ & $\mathcal{A}_{21}(\log r + c_{21})$ & $(-2.859 \pm 0.117) \times 10^{-3}$ & $-2.711 \times 10^{-3}$ & $1.055$ \\[2pt]
  $C_{3,1}$ & $\mathcal{A}_{31}(\log r + c_{31})$ & $(-5.612 \pm 0.071) \times 10^{-3}$ & $-5.769 \times 10^{-3}$ & $0.973$ \\[2pt]
  $C_{4,1}$\xrowht{12pt} & $\mathcal{A}_{41}(\log r + c_{41})$ & $(\phantom{-}8.519 \pm 0.203) \times 10^{-3}$ & $\phantom{-}8.480 \times 10^{-3}$ & $1.005$ \\
  \midrule
  $C_{2,2}$ & $\mathcal{A}_{22}\log^2r+ \mathcal{B}_{22}\log r + \mathcal{D}_{22} $ & $(-2.404 \pm 0.211) \times 10^{-3}$ & $-2.711 \times 10^{-3}$ & $0.887$ \\
  \bottomrule
\end{tabular}
\caption{Leading amplitudes $\mathcal{A}_{ab}$ for the square lattice two-point correlations ($L=8000\,\ell$). $C_{4,1}$ is reconstructed via the sum rule $C_{4,1}=-(C_{1,1}+C_{2,1}+C_{3,1})$. Data are obtained by sampling over $n_{\text{total}}\approx10^6$ configurations. Fit windows: $C_{a,1}$ on $r\in[6,20]$, $C_{2,2}$ on $r\in[1,20]$. The quoted uncertainty is the standard error of the fitted coefficient (and of the mean for $C_{1,1}$). Theory: $\mathcal{A}_{11}=\mathcal{A}_{21}=\mathcal{A}_{22}=-P_1^2/2$, $\mathcal{A}_{31}=-(\pi-2)(8-\pi)/\pi^6$, $\mathcal{A}_{41}=(\pi+4)(\pi-2)/\pi^6$.}
\label{tab:square_fits}
\end{table}

The sum rule $\sum_a C_{a,1}(r)=0$, which follows from $\sum_a \delta_{h(i),a}=1$, allows us to infer $C_{4,1}(r)$ from the three measured $C_{a,1}$ without computing it directly:
\begin{equation}
  C_{4,1}(r) = -\Bigl[C_{1,1}(r) + C_{2,1}(r) + C_{3,1}(r)\Bigr].
  \label{eq:C14_sumrule}
\end{equation}
The asymptotic slope of $C_{4,1}$ must be positive: $C_{1,1}$ is asymptotically constant and negative, while $C_{2,1}$ and $C_{3,1}$ have negative slopes (see Table~\ref{tab:square_fits}), so their sum is asymptotically negative, forcing $C_{4,1}$ to have a positive slope. Physically, this reflects the FSC constraint---height~1 is minimally stable and suppresses nearby low heights ($h=1,2,3$), so the maximally stable height~4 must be \emph{enhanced} near a height~1 site. A linear fit to the reconstructed $r^4 C_{4,1}(r)$ over $r\in[6,20]$ gives $\mathcal{A}_{41}^{\text{fit}}=(8.519\pm0.203)\times 10^{-3}$. The exact prediction from the field decomposition yields $\mathcal{A}_{41}^{\text{theory}} = -\alpha_4\beta_1 P_1^2/2= (\pi+4)(\pi-2)/\pi^6 \approx 8.48\times 10^{-3}$, in agreement with the fitted value to within $0.5\%$.

The sub-leading shift constants are also reasonably well reproduced: for $C_{2,1}$, $c_{21}^{\text{fit}} = 1.485$ vs.\ the exact $c_{21} = 1.745$ (deviation $-14.9\%$); for $C_{3,1}$, $c_{31}^{\text{fit}} = -0.797$ vs.\ $c_{31} = -0.837$ (deviation $-4.8\%$).

%%%%%%%%%%%%%%%%%%%%%%%%%%%%
\subsection{Honeycomb lattice}
\label{sec:honeycomb}
%%%%%%%%%%%%%%%%%%%%%%%%%%%%

The honeycomb lattice is a bipartite lattice with two sublattices (A, B) per unit cell and coordination number~$z=3$, studied here on a hexagonal domain of radius~$L$ [see Figure~\ref{fig:lattices}(b)]. In contrast to the square and kagome cases, the allowed heights are $h\in\{1,2,3\}$. Exact bulk height probabilities are known, and the leading asymptotic coefficient of $C_{1,1}(r)$ has also been derived analytically~\cite{PonceletRuelle18,Honeycomb10}. For the mixed correlators $C_{2,1}$ and $C_{3,1}$, however, we are not aware of closed-form honeycomb predictions for the amplitudes, so our numerical analysis in that sector is limited to the functional form and fitted slopes.

%%%%%%%%%%%%%%%%%%%%%%%%%%%%
\subsubsection{Height probabilities}
%%%%%%%%%%%%%%%%%%%%%%%%%%%%

For the full honeycomb lattice the one-site probabilities are known exactly~\cite{PonceletRuelle18}:
\begin{equation}
  P_1 = \frac{1}{12}, \qquad
  P_2 = \frac{7}{24}, \qquad
  P_3 = \frac{5}{8}.
\end{equation}
Table~\ref{tab:honeycomb_heights} compares these values with the measured probabilities obtained on a hexagonal domain of radius $L=6000$ using a merged data set with $n_{\text{total}}=8\times 10^5$ Wilson samples. The two sublattices are statistically indistinguishable within our precision, and the average agrees with the exact values at the $10^{-6}$ level.
\begin{table}[h]
\centering
\begin{tabular}{lcccc}
  \toprule
  height & analytic prediction & sub A & sub B & average \\
  \midrule
  $P_1$ & $1/12 = 0.083333333$ & $0.083333451$ & $0.083333365$ & $0.083333408$ \\
  $P_2$ & $7/24 = 0.291666667$ & $0.291667199$ & $0.291667360$ & $0.291667279$ \\
  $P_3$ & $\phantom{o}5/8 = 0.625000000$ & $0.624999350$ & $0.624999275$ & $0.624999312$ \\
  \bottomrule
\end{tabular}
\caption{Honeycomb lattice height probabilities ($L=6000\sqrt{3}\,\ell$, $n\approx8\times10^5$).  Exact: $P_1=1/12$, $P_2=7/24$, $P_3=5/8$~\cite{PonceletRuelle18}.  Sub A/sub B are the per-sublattice values from the nn-persub merge; the two sublattices agree within statistical uncertainty.}
\label{tab:honeycomb_heights}
\end{table}

%%%%%%%%%%%%%%%%%%%%%%%%%%%%
\subsubsection{Two-point correlation functions}
\label{sec:honeycomb_corr}
%%%%%%%%%%%%%%%%%%%%%%%%%%%%

\begin{figure}[t]
\centering
\includegraphics[width=0.95\textwidth]{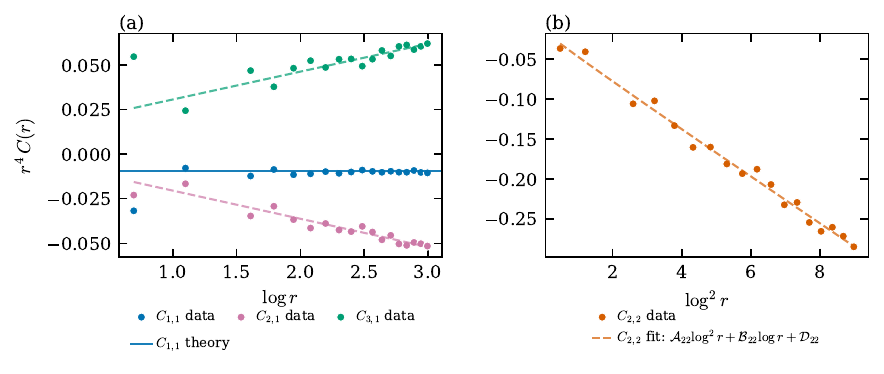}
\caption{Honeycomb lattice rescaled correlations ($L=6000\sqrt{3}\,\ell$, $n_{\text{total}}\approx 10^6$), averaged over sublattices A and B; distances in nn (bond-length $\ell$) units.  (a)~$r^4C_{1,1}$ (blue), $r^4C_{2,1}$ (pink, linear fit), $r^4C_{3,1}$ (green, linear fit) vs $\log r$ ($r\in[6,20]$); solid line: exact $-9/(96\pi^2)$ in $\ell$-units.  (b)~$r^4C_{2,2}$ (orange) vs $\log^2 r$ with three-parameter fit ($r\in[2,20]$).}
\label{fig:honeycomb_logr}
\end{figure}

Having verified the universal exponent $2\Delta = 4$ on the honeycomb lattice in Section~\ref{sec:universal}, we now fix the prefactor $r^4$ and study the rescaled honeycomb correlations. Figure~\ref{fig:honeycomb_logr} presents the numerical data for $r^4 C_{1,1}(r)$, $r^4 C_{2,1}(r)$, $r^4 C_{3,1}(r)$ and $r^4 C_{2,2}(r)$. For $C_{1,1}$ the exact leading coefficient was determined in Reference~\cite{Honeycomb10}. Since we measure distances in units of the bond length $\ell$ we have to rescale by $(a/\ell)^4=9$, yielding
\begin{equation}
  \mathcal{A}_{11}^{\mathrm{hon}} = -\frac{9}{96\pi^2}  \approx -9.499 \times 10^{-3},
\end{equation}
For $C_{2,1}$, $C_{3,1}$ and $C_{2,2}$, we are not aware of exact bulk results on the honeycomb lattice, so we only quote the fitted numerical values.

\begin{table}[h]
\centering
\begin{tabular}{lllc}
  \toprule
  $r^4\times$ & fit form & fit range & $\mathcal{A}_{ab}^{\text{fit}}$ \\
  \midrule
  $C_{1,1}$ & $\mathcal{A}_{11}$ & $r\in[6,20]$  & $ (-1.000 \pm 0.020) \times 10^{-2}$ \\[3pt]
  $C_{2,1}$ & $\mathcal{A}_{21}\log r + \mathrm{const.}$ & $r\in[6,20]$ & $(-1.579 \pm 0.157) \times 10^{-2}$ \\[3pt]
  $C_{3,1}$ & $\mathcal{A}_{31}\log r + \mathrm{const.}$ & $r\in[6,20]$ & $(\phantom{-}1.563 \pm 0.206) \times 10^{-2}$ \\[3pt]
  \midrule
  $C_{2,2}$ & $\mathcal{A}_{22}\log^2 r + \mathcal{B}_{22}\log r + \mathcal{D}_{22}$ & $r\in[2,20]$ & $(-2.824 \pm 0.486) \times 10^{-2}$ \\
  \bottomrule
\end{tabular}
\caption{Fit results for the honeycomb lattice rescaled correlations $r^4C_{a,b}(r)$ ($\ell$ units).  An exact bulk value is available only for $\mathcal{A}_{11}^{\mathrm{hon}}=-9/(96\pi^2)\approx -9.499\times 10^{-3}$ (in $\ell$-units), implying $\mathcal{A}_{11}^{\text{fit}}/\mathcal{A}_{11}^{\mathrm{hon}}=1.053$. The other channels are numerical fits to the sublattice-averaged curves. The quoted uncertainties are standard errors of the fitted leading coefficients (and of the mean for the constant $C_{1,1}$ fit).  Fit windows: $C_{a,1}$ on $r\in[6,20]$, $C_{2,2}$ on $r\in[2,20]$.}
\label{tab:honeycomb_amplitudes}
\end{table}

The fitted constant for $C_{1,1}$ differs from the exact value by about $5.3\%$, which is reasonable for the current lattice size and fit range. The mean inter-sublattice standard deviations over the fit windows are $1.20\times 10^{-5}$ for $C_{1,1}$, $1.43\times 10^{-5}$ for $C_{2,1}$, $7.73\times 10^{-6}$ for $C_{3,1}$, and $1.28\times 10^{-5}$ for $C_{2,2}$. We note that as the fit range for $C_{2,2}$ we choose $r\in[2,20]$ (instead of $r_\min=1$), since the same-sublattice data do not contain pairs at $r=1$. 

The mixed correlators are fully consistent with the expected linear dependence on $\log r$: the $C_{2,1}$ slope is negative, the $C_{3,1}$ slope is positive, and the two magnitudes are very similar. The positive $C_{3,1}$ slope follows from the sum rule $C_{3,1}=-(C_{1,1}+C_{2,1})$ (since the honeycomb lattice has only three heights) and reflects the same FSC mechanism as on the square lattice: the maximally stable height~3 is enhanced near a height~1 site. Finally, $C_{2,2}$ is well described by a quadratic polynomial in $\log r$, again matching the generic LCFT structure even though no exact honeycomb coefficients are currently available. We will further analyse the amplitudes in Section~\ref{sec:CFTcheck} below.

%%%%%%%%%%%%%%%%%%%%%%%%%%%%
\subsection{Kagome lattice}
\label{sec:kagome}
%%%%%%%%%%%%%%%%%%%%%%%%%%%%

The kagome lattice is a lattice of corner-sharing triangles with three sublattices (A, B, C) per unit cell, studied on a hexagonal domain of radius~$L$ [see Figure~\ref{fig:lattices}(c) for the geometry and the definition of~$L$]. Despite having the same coordination number $z=4$ as the square lattice, its non-Bravais structure leads to distinct height statistics and correlation functions. While the height probabilities $P_a$ are not known analytically, the sandpile density $\zeta^{\text{kag}} = \sum_a aP_a=19/6 \approx 3.1\overline{6}$ has been computed exactly~\cite{KasselWilson16}.

%%%%%%%%%%%%%%%%%%%%%%%%%%%%
\subsubsection{Analytic height-1 probability}\label{sec:kagomeP1analytical}
%%%%%%%%%%%%%%%%%%%%%%%%%%%%

\begin{figure}[t]
\centering
\includegraphics[width=\textwidth]{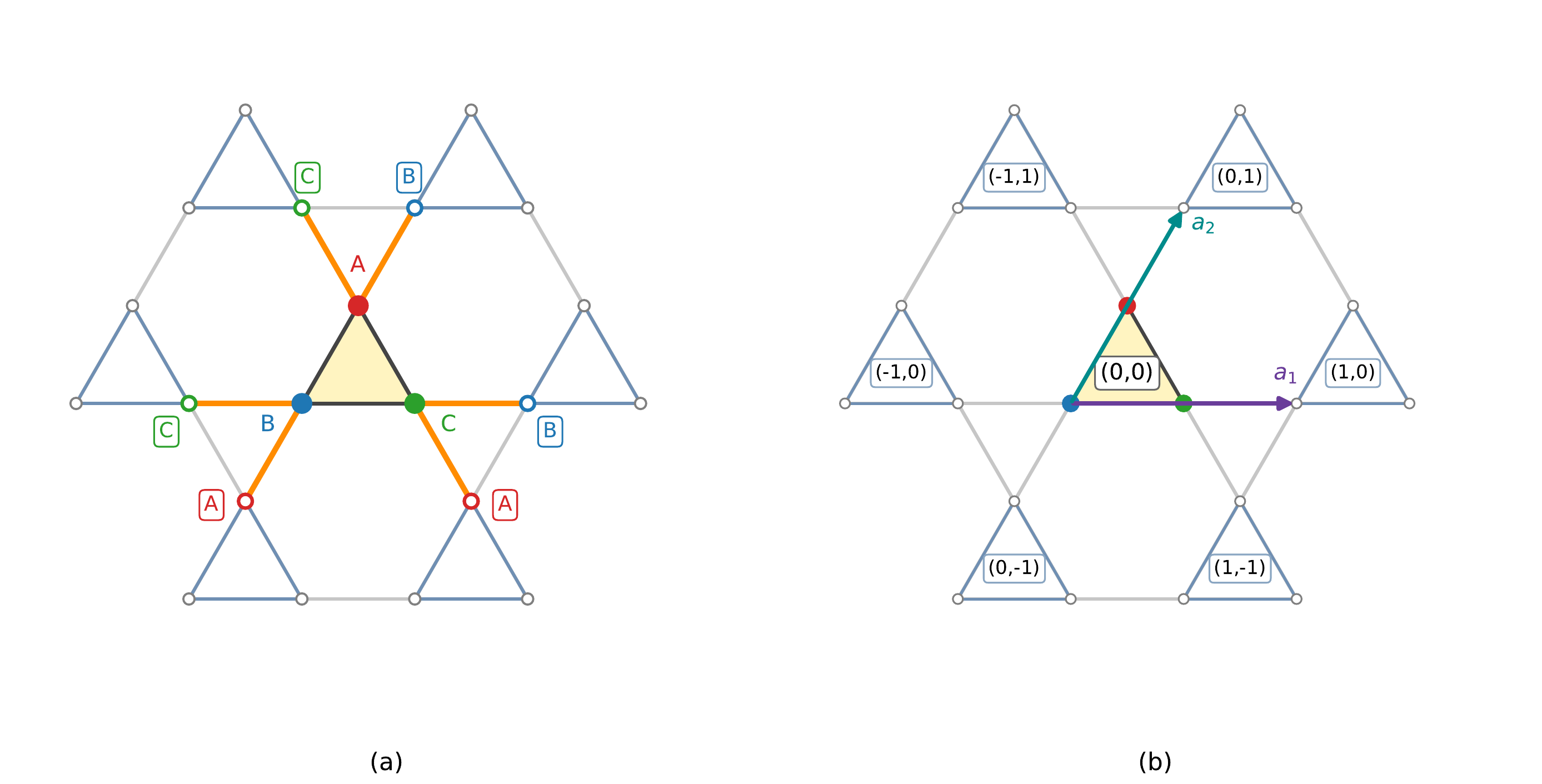}
\caption{Kagome lattice geometry used in the analytic discussion of bulk height probabilities. (a)~Reference unit cell $(0,0)$ with sublattices A, B, C, together with the six nearest sites connected to it. (b)~The same local kagome geometry with the primitive lattice vectors $a_1$ and $a_2$, chosen parallel to the bonds BC and BA, respectively.}
\label{fig:kagome_unit_cell_zoom}
\end{figure}
\begin{figure}[t]
\centering
\includegraphics[width=0.45\textwidth]{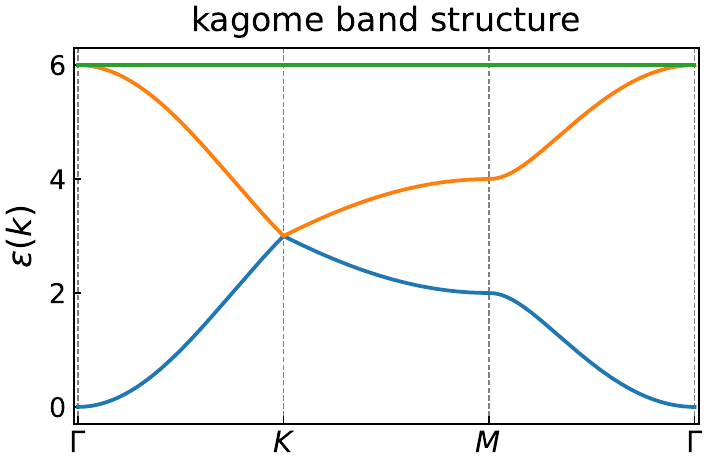}
\caption{Band structure of the Bloch Laplacian associated with the kagome unit-cell convention used in Figure~\ref{fig:kagome_unit_cell_zoom}. The presence of a flat band together with two dispersive branches provides a simple consistency check that the block matrices above indeed describe the kagome lattice.}
\label{fig:kagome_bands_appendix}
\end{figure}

Here we derive the height-1 probability $P_1$. Using the six highlighted links in Figure~\ref{fig:kagome_unit_cell_zoom}(a), the nearest-neighbour connectivity can be organized into $3\times 3$ block matrices labelled by the displacement between unit cells. In the sublattice basis (A,B,C) of the reference cell $(0,0)$, the intracell block is
\begin{equation}
  a(0,0)=
  \begin{pmatrix}
    0 & 1 & 1 \\
    1 & 0 & 1 \\
    1 & 1 & 0
  \end{pmatrix},
\end{equation}
while the three independent intercell blocks are
\begin{equation}
  a(1,0)=
  \begin{pmatrix}
    0 & 0 & 0 \\
    0 & 0 & 0 \\
    0 & 1 & 0
  \end{pmatrix},
  \qquad
  a(0,1)=
  \begin{pmatrix}
    0 & 1 & 0 \\
    0 & 0 & 0 \\
    0 & 0 & 0
  \end{pmatrix},
  \qquad
  a(1,-1)=
  \begin{pmatrix}
    0 & 0 & 0 \\
    0 & 0 & 0 \\
    1 & 0 & 0
  \end{pmatrix}.
\end{equation}
The opposite displacements are fixed by Hermiticity,
\begin{equation}
  a(-1,0)=a(1,0)^T,\qquad
  a(0,-1)=a(0,1)^T,\qquad
  a(-1,1)=a(1,-1)^T.
\end{equation}
The corresponding Bloch Laplacian is therefore
\begin{align}
  \widetilde{\Delta}(\theta_1,\theta_2)
  ={}& 4I
  - a(0,0)
  - a(1,0)e^{\ii\theta_1}
  - a(0,1)e^{\ii\theta_2}
  - a(1,-1)e^{\ii(\theta_1-\theta_2)}
  \nonumber\\
  &- a(-1,0)e^{-\ii\theta_1}
  - a(0,-1)e^{-\ii\theta_2}
  - a(-1,1)e^{-\ii(\theta_1-\theta_2)}.
\end{align}
Here $\theta_1,\theta_2$  are the Bloch momenta conjugate to the primitive vectors $a_1,a_2$, so that a unit-cell displacement $(n_1,n_2)$ carries the phase $e^{\ii(n_1\theta_1+n_2\theta_2)}$. Assuming an $N_1\times N_2$ array of unit cells periodic in both coordinates, the momenta are discrete, $\theta_j=2\pi k_j/N_j$ with $0\le k_j\le N_j-1$, and in the infinite-lattice limit $N_1,N_2\to\infty$ momentum sums become continuous integrals over the Brillouin zone $[0,2\pi)^2$. The resulting translation-invariant Green function is a bulk quantity, which we compare with the bulk of the finite domains used in our simulations (radius-$L$ hexagons with wired boundaries, Figure~\ref{fig:lattices}), the two agreeing up to finite-size corrections. The corresponding Bloch matrix exhibits the characteristic kagome band structure, with one flat band and two dispersive bands. For the present choice of unit cell and primitive vectors, the bands along a standard high-symmetry path are shown in Figure~\ref{fig:kagome_bands_appendix}.

To compute the height-1 probability, we follow the same defect-matrix logic as on the honeycomb lattice~\cite{Honeycomb10}. We choose the reference site to be $A(0,0)$, ie, the red site in the centre triangle in Figure~\ref{fig:kagome_unit_cell_zoom}(b), and remove three of its four incident bonds, namely the links to $C(-1,1)$, $B(0,0)$, and $C(0,0)$, leaving only the bond to $B(0,1)$. With the site ordering
\begin{equation}
  \bigl\{A(0,0),\; C(-1,1),\; B(0,0),\; C(0,0)\bigr\},
\end{equation}
the defect matrix is supported only on these four sites and reads 
\begin{equation}
  D=
  \begin{pmatrix}
    3 & -1 & -1 & -1 \\
    -1 & 1 & 0 & 0 \\
    -1 & 0 & 1 & 0 \\
    -1 & 0 & 0 & 1
  \end{pmatrix}.
\end{equation}
The corresponding restriction of the infinite-volume Green function to the same four-site support is
\begin{equation}
  G=
  \begin{pmatrix}
    G_{AA}(0,0) & G_{AC}(-1,1) & G_{AB}(0,0) & G_{AC}(0,0) \\
    G_{AC}(-1,1) & G_{CC}(0,0) & G_{BC}(-1,1) & G_{CC}(-1,1) \\
    G_{AB}(0,0) & G_{BC}(-1,1) & G_{BB}(0,0) & G_{BC}(0,0) \\
    G_{AC}(0,0) & G_{CC}(-1,1) & G_{BC}(0,0) & G_{CC}(0,0)
  \end{pmatrix},
\end{equation}
where translation invariance implies that $G$ depends only the on relative displacement, while the symmetry of the toppling matrix gives
\begin{equation}
  G_{\alpha\beta}(n_1,n_2)=G_{\beta\alpha}(-n_1,-n_2).
\end{equation}
If $\Delta$ denotes the infinite-volume toppling matrix and $\Delta'=\Delta-D$ the defect-modified one, then
\begin{equation}
  P_1^{\mathrm{kag}}
  =
  \frac{\det(\Delta-D)}{\det\Delta}
  =
  \det\!\left(I-\Delta^{-1}D\right),
\end{equation}
where only the restriction of $G=\Delta^{-1}$ to the support of $D$ is needed. Since each row and each column of $D$ sums to zero, one may subtract the same constant from every entry of this restricted Green matrix without changing the determinant. It is therefore convenient to introduce the relative local Green matrix
\begin{equation}
  G_{\mathrm{loc}}
  :=
  G-G_{AA}(0,0)\,
  \begin{pmatrix}
    1&1&1&1\\
    1&1&1&1\\
    1&1&1&1\\
    1&1&1&1
  \end{pmatrix},
\end{equation}
for which
\begin{equation}
  P_1^{\mathrm{kag}}=\det(I-G_{\mathrm{loc}}D).
  \label{eq:P1kagomeanalytic}
\end{equation}
Using bulk symmetry and the relative Green-function values derived in Appendix~\ref{app:kagome_green},
\begin{equation}
  G_{AB}(0,0)-G_{AA}(0,0)
  =
  G_{AC}(0,0)-G_{AA}(0,0)
  =
  G_{BC}(0,0)-G_{AA}(0,0)
  =
  -\frac14,
\end{equation}
\begin{equation}
  G_{BC}(-1,1)-G_{AA}(0,0)
  =
  -\frac{5}{18}-\frac{\sqrt3}{6\pi},
  \quad
  G_{CC}(-1,1)-G_{AA}(0,0)
  =
  -\frac{17}{36}+\frac{\sqrt3}{6\pi},
\end{equation}
we obtain
\begin{equation}
  G_{\mathrm{loc}}
  =
  \begin{pmatrix}
    0 & -\frac14 & -\frac14 & -\frac14 \\
    -\frac14 & 0 & -\frac{5}{18}-\frac{\sqrt3}{6\pi} & -\frac{17}{36}+\frac{\sqrt3}{6\pi} \\
    -\frac14 & -\frac{5}{18}-\frac{\sqrt3}{6\pi} & 0 & -\frac14 \\
    -\frac14 & -\frac{17}{36}+\frac{\sqrt3}{6\pi} & -\frac14 & 0
  \end{pmatrix}.
\end{equation}
Thus the height-1 probability is reduced to the determinant of the finite $4\times4$ matrix \eqref{eq:P1kagomeanalytic}, with the final exact value
\begin{equation}
  P_1^{\mathrm{kag}}
  =
  \frac{85}{1296}
  + \frac{7\sqrt3}{432\pi}
  - \frac{1}{24\pi^2}
  \approx 0.070298273832694,
  \label{eq:P1kagfinal}
\end{equation}
which is the quantity most directly needed for the asymptotic discussion below. By symmetry, this value is identical on all three kagome sublattices A, B and C. The corresponding exact expressions for $P_2$, $P_3$ and $P_4$ can in principle be obtained in the same framework, but their analytic forms are considerably more involved and will not be derived here.

%%%%%%%%%%%%%%%%%%%%%%%%%%%%
\subsubsection{Numerical height probabilities}
%%%%%%%%%%%%%%%%%%%%%%%%%%%%

Table~\ref{tab:kagome_heights} shows the kagome height probabilities measured on a hexagonal domain with linear size $L=12000\,\ell$ using $n_{\text{total}}=1.2\times10^6$ samples. The three sublattices have nearly identical distributions within statistical uncertainty, in agreement with the analytic prediction. We note that the height-1 probability $P_1^{\text{kag}} \approx 0.0703$ is slightly smaller than for the square lattice ($P_1^{\text{sq}} \approx 0.0736$). We also quote the sandpile density $\zeta^{\text{kag}}$. 

\begin{table}[h]
\centering
\begin{tabular}{lcccc}
  \toprule
  height & sub A & sub B & sub C & average \\
  \midrule
  $P_1$ & $0.07029836$ & $0.07029838$ & $0.07029830$ & $0.070298346$ \\
  $P_2$ & $0.15087392$ & $0.15087401$ & $0.15087389$ & $0.150873941$ \\
  $P_3$ & $0.32069270$ & $0.32069255$ & $0.32069259$ & $0.320692615$ \\
  $P_4$ & $0.45813502$ & $0.45813506$ & $0.45813521$ & $0.458135098$ \\
  $P_1^{\mathrm{exact}}$ & \multicolumn{4}{c}{$0.070298273832694$} \\
  rel.\ error in $P_1$ & \multicolumn{4}{c}{$1.028 \times 10^{-6}$} \\
  \midrule
  $\zeta = \sum a\,P_a$ & $3.166664$ & $3.166664$ & $3.166665$ & $3.166664$ \\
  $\zeta_{\text{exact}}$ & \multicolumn{4}{c}{$19/6 \approx 3.166667$} \\
   \bottomrule
\end{tabular}
\caption{Kagome lattice height probabilities by sublattice ($L=12000\,\ell$, $m=60$, $n_{\text{total}}=1.2\times10^6$). The three sublattices are statistically indistinguishable. The quoted relative error for $P_1$ is computed from the unrounded numerical average $0.0702983461213532$ and the analytical value derived in Section~\ref{sec:kagome}. The measured density $\zeta$ agrees with the exact value $19/6$~\cite{KasselWilson16} to within $0.001\%$.}
\label{tab:kagome_heights}
\end{table}

%%%%%%%%%%%%%%%%%%%%%%%%%%%%
\subsubsection{Two-point correlation functions}
\label{sec:kagome_corr}
%%%%%%%%%%%%%%%%%%%%%%%%%%%%

\begin{figure}[t]
\centering
\includegraphics[width=0.95\textwidth]{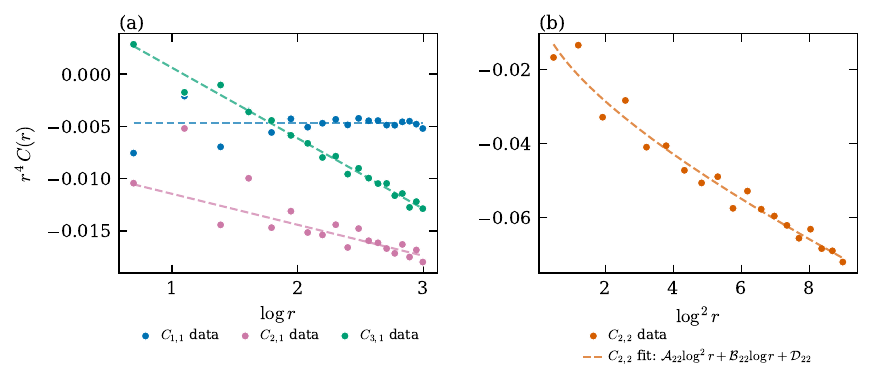}
\caption{Kagome lattice rescaled correlations ($L=12000\,\ell$, $n_{\text{total}}=1.2\times10^6$), averaged over sublattices A, B, C; distances in nn (bond-length $\ell$) units.  (a)~$r^4C_{1,1}$ (blue, constant fit), $r^4C_{2,1}$ (pink, linear fit), $r^4C_{3,1}$ (green, linear fit) vs $\log r$ ($r\in[6,20]$).  (b)~$r^4C_{2,2}$ (orange) vs $\log^2 r$ with three-parameter fit ($r\in[2,20]$).}
\label{fig:kagome_logr}
\end{figure}

We now determine the lattice-dependent coefficients for the correlation functions on the kagome lattice. Figure~\ref{fig:kagome_logr} presents the rescaled kagome correlations. Panel~(a) shows the sublattice-averaged $r^4\,C(r)$ vs.\ $\log r$ for $C_{1,1}$ (constant, Case~1), $C_{2,1}$ and $C_{3,1}$ (linear, Case~2), while panel~(b) displays the averaged $r^4\,C_{2,2}(r)$ vs.\ $\log^2 r$ (quadratic, Case~3). The residual inter-sublattice spread is quoted numerically in the text below, while colours distinguish the correlation type. Table~\ref{tab:kagome_amplitudes} summarises the fit results. We compare each correlation with the corresponding asymptotic form from Section~\ref{sec:CFT}.
\begin{table}[t]
\centering
\begin{tabular}{lllc}
  \toprule
  $r^4\times$ & fit form & fit range & $\mathcal{A}_{ab}^{\text{fit}}$ \\
  \midrule
  $C_{1,1}$ & $\mathcal{A}_{11}$ & $r\in[6,20]$ & $(-4.715 \pm 0.097) \times 10^{-3}$ \\[3pt]
  $C_{2,1}$ & $\mathcal{A}_{21}(\log r + c_{21})$ & $r\in[6,20]$ & $(-2.967 \pm 0.516) \times 10^{-3}$ \\[3pt]
  $C_{3,1}$ & $\mathcal{A}_{31}(\log r + c_{31})$ & $r\in[6,20]$ & $(-6.758 \pm 0.282) \times 10^{-3}$ \\[3pt]
  $C_{4,1}$\xrowht{12pt} & $\mathcal{A}_{41}(\log r + c_{41})$ (reconstructed) & $r\in[6,20]$ & $(\phantom{-}9.583 \pm 0.892) \times 10^{-3}$ \\[3pt]
  \midrule
  $C_{2,2}$ & $\mathcal{A}_{22}\log^2 r + \mathcal{B}_{22}\log r + \mathcal{D}_{22}$ & $r\in[2,20]$ & $(-2.355 \pm 1.865) \times 10^{-3}$ \\
  \bottomrule
\end{tabular}
\caption{Fit results for the kagome lattice rescaled correlations $r^4C_{a,b}(r)$ ($\ell$ units). $C_{4,1}$ is first reconstructed pointwise from the sum rule $C_{4,1}=-(C_{1,1}+C_{2,1}+C_{3,1})$ and then fitted to the same linear form. There are no closed-form bulk amplitudes; values are numerical fits to sublattice-averaged curves. The quoted uncertainties are standard errors of the fitted leading coefficients (and of the mean for the constant $C_{1,1}$ fit).}
\label{tab:kagome_amplitudes}
\end{table}

We note that the mean inter-sublattice standard deviations of the rescaled quantities $r^4C$ over the fit windows are $1.00 \times 10^{-4}$ for $C_{1,1}$, $9.32 \times 10^{-5}$ for $C_{2,1}$, $1.37 \times 10^{-4}$ for $C_{3,1}$, and $1.73 \times 10^{-4}$ for $C_{2,2}$. These values remain small throughout, confirming that the non-Bravais sublattice structure does not produce any visible splitting between the three kagome sublattices at the current level of precision. The sum rule $\sum_a C_{a,1}=0$ again allows us to reconstruct $C_{4,1}$, whose fitted slope $\mathcal{A}_{41} \approx +9.58\times 10^{-3}$ (see Table~\ref{tab:kagome_amplitudes}) is positive, consistent with the same FSC mechanism discussed for the square and honeycomb lattices. The comparatively large standard error of $\mathcal{A}_{22}$ reflects the sensitivity of the three-parameter quadratic fit to the available distance range and to statistical fluctuations in the large-$r$ tail. A more precise determination of this leading coefficient would require a larger lattice and/or more independent samples than are currently computationally accessible. Nevertheless, the observed curvature is consistent with the predicted $\log^2 r$ dependence, although the present data do not permit a high-precision estimate of $\mathcal{A}_{22}$.

%%%%%%%%%%%%%%%%%%%%%%%%%%%%
\subsection{Numerical determination of LCFT amplitudes}
\label{sec:CFTcheck}
%%%%%%%%%%%%%%%%%%%%%%%%%%%%

In this section we aim to determine the expansion parameters $\alpha_a$ and $\beta_a$ in the decomposition \eqref{eq:field_decomp} beyond the square lattice. By construction $\alpha_1=\beta_2=0$. The numerical results for the leading amplitudes $\mathcal{A}_{ab}$ given in Tables~\ref{tab:honeycomb_amplitudes} and~\ref{tab:kagome_amplitudes} allow us to obtain estimates for the some of the remaining expansion parameters $\alpha_a$ and $\beta_a$.

To keep the extraction uniform across the two non-square lattices, we use the numerical $P_1$ in Tables~\ref{tab:honeycomb_heights} and~\ref{tab:kagome_heights} together with the fitted leading amplitudes $\mathcal{A}_{ab}$ in Tables~\ref{tab:honeycomb_amplitudes} and~\ref{tab:kagome_amplitudes}, respectively. Comparing the leading behaviour of $C_{1,1}$ with \eqref{eq:leadingCF} gives
\begin{equation}
\beta_1=\sqrt{\frac{\mathcal{A}_{11}^{\mathrm{fit}}}{A}},\qquad A=-\frac{P_1^2}{2},
\end{equation}
where we pick the positive root for $\beta_1$. For the honeycomb lattice this parameter is also known exactly: using $P_1=1/12$ and $\mathcal{A}_{11}=-9/(96\pi^2)$ in $\ell$-units gives
\begin{equation}
\beta_1^{\mathrm{exact}}=\sqrt{\frac{-9/(96\pi^2)}{-(1/12)^2/2}}=\frac{3\sqrt{3}}{\pi}\approx1.654.
\end{equation}
The numerical estimate is about $2.6\%$ larger than this exact value. The parameters $\alpha_2$ and $\alpha_3$ can now be obtained from $\mathcal{A}_{a1}=\alpha_a\beta_1A$. For the kagome lattice we use the same numerical procedure for all parameters $\beta_1$ and $\alpha_a$, $a=2,3,4$. In particular, $\alpha_4$ is obtained from the reconstructed $\mathcal{A}_{41}$ in Table~\ref{tab:kagome_amplitudes}. The quoted uncertainties are propagated from the fitted amplitudes; the much smaller uncertainty of the numerical $P_1$ is not included. We summarise the obtained estimates in Table~\ref{tab:expansion parameters}.
\begin{table}[t]
\centering
\begin{tabular}{lccc}
  \toprule
  parameter & square (exact) & honeycomb & kagome \\
  \midrule
  $\beta_1$ & $\phantom{-}1$ & $\phantom{-}1.697 \pm 0.017$ & $\phantom{-}1.381 \pm 0.014$ \\[3pt]
  $\alpha_2$ & $\phantom{-}1$ & $\phantom{-}2.679 \pm 0.268$ & $\phantom{-}0.869 \pm 0.151$ \\[3pt]
  $\alpha_3$ & $\phantom{-}2.128$ & $-2.652 \pm 0.350$ & $\phantom{-}1.980 \pm 0.085$ \\[3pt]
  $\alpha_4$ & $-3.128$ & - & $-2.808 \pm 0.263$ \\
  \bottomrule
\end{tabular}
\caption{Numerical estimates of $\beta_1$ and the available coefficients $\alpha_a$ in the field decomposition \eqref{eq:field_decomp} for the honeycomb and kagome lattices. The exact honeycomb value of $\beta_1$ is given by~\cite{Honeycomb10} $\beta_1=3\sqrt{3}/\pi\approx1.654$. We also state the exact results \eqref{eq:alpha_square} and \eqref{eq:beta_square} for the square lattice for comparison.}
\label{tab:expansion parameters}
\end{table}

\begin{figure}[t]
\centering
\includegraphics[width=\textwidth]{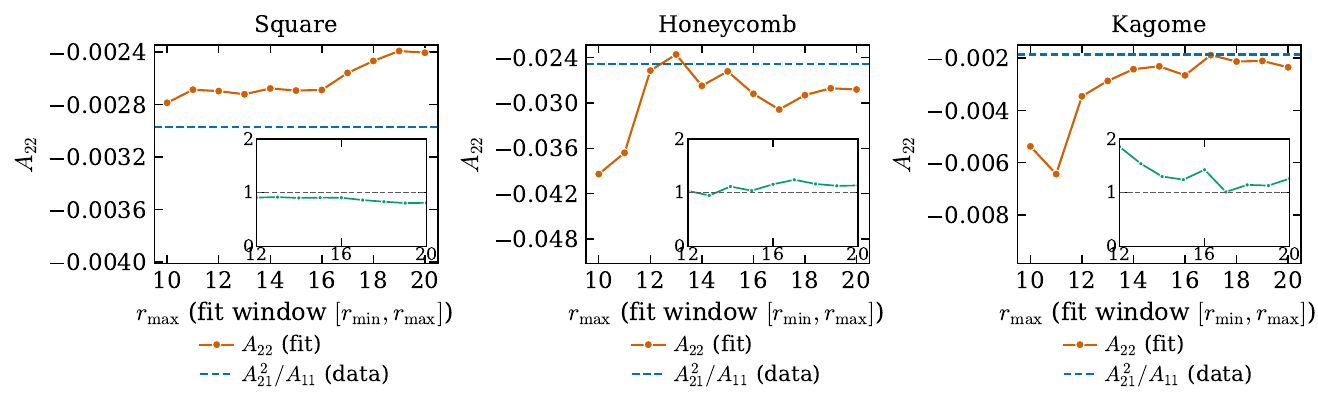}
\caption{Stability of the $\mathcal{A}_{22}$ amplitude from the three-parameter quadratic fit as function of the upper fit bound $r_\max=10,\ldots,20$.  The $\mathcal{A}_{22}$ fits use the window $r\in[r_{\min},r_\max]$ in nearest-neighbour (bond-length $\ell$) units, with $r_{\min}=1$ for the square lattice and $r_{\min}=2$ for the honeycomb and kagome lattices.  Dashed line: the prediction $\mathcal{A}_{21}^2/\mathcal{A}_{11}$ from the $C_{a,1}$ data fits on $r\in[6,20]$ for all three lattices. Insets show the ratio \eqref{eq:CFTratio} of the fitted $\mathcal{A}_{22}$ to the corresponding dashed reference value for $r_\max=12,\ldots,20$, using the common vertical range $0\le R\le2$.}
\label{fig:a22_scan}
\end{figure}

Having obtained the amplitudes, the LCFT predictions allow us an additional consistency check on our results. Specifically, the leading amplitude for the correlation functions $C_{2,2}$ has to satisfy
\begin{equation}
\mathcal{A}_{22}=\alpha_2^2A\stackrel{!}{=}\frac{\mathcal{A}_{21}^2}{\mathcal{A}_{11}}.
\end{equation}
In Figure~\ref{fig:a22_scan} we show the values of $\mathcal{A}_{22}$ obtained by fitting the $C_{2,2}$ correlation functions on the window $r\in[r_\min,r_\max]$ as a function of the upper fit bound $r_\max=10,\ldots,20$. We use $r_{\min}=1$ for the square lattice and $r_{\min}=2$ for the honeycomb and kagome lattices. We also include the predicted values $\mathcal{A}_{21}^2/\mathcal{A}_{11}$ obtained from data fits to $C_{2,1}$ and $C_{1,1}$ on the fixed window $r\in[6,20]$ for all three lattices. The insets display the ratio
\begin{equation}
R(r_\max)=\frac{\mathcal{A}_{22}^{\mathrm{fit}}(r_\max)}
{\mathcal{A}_{21}^2/\mathcal{A}_{11}},
\label{eq:CFTratio}
\end{equation}
for which the LCFT relation above predicts $R=1$. On the square lattice, $R\simeq0.90$--$0.92$ for $r_\max=12$--$16$, before decreasing to about $0.81$ at $r_\max=20$. For the honeycomb lattice, the range $r_\max=12$--$20$ gives $R\simeq0.95$--$1.24$. The kagome result is more sensitive to the fit window, but approaches unity for the larger intermediate windows, with $R\simeq1.01$ at $r_\max=17$ and $R\simeq1.13$--$1.26$ for $r_\max=18$--$20$. Thus, the data support the relation $\mathcal{A}_{22}=\mathcal{A}_{21}^2/\mathcal{A}_{11}$ to varying degrees on all three lattices, although the residual window dependence, especially for kagome, prevents a high-precision test. We nevertheless conclude that our numerical data for the correlation functions are consistent with the asymptotic forms predicted by the LCFT framework.

%%%%%%%%%%%%%%%%%%%%%%%%%%%%
\section{Discussion and conclusion}
\label{sec:discussion}
%%%%%%%%%%%%%%%%%%%%%%%%%%%%

We have presented a numerical study of height-height correlations in the Abelian sandpile model on the square, honeycomb and kagome lattices, using Wilson's algorithm for uniform spanning tree sampling combined with a FFT-based correlation computation.

\paragraph{Universality of the scaling exponent.}
On all three lattices a free power-law fit to $-C_{1,1}(r) \sim r^{-2\Delta}$ yields $2\Delta \approx 4$, confirming that all three lattices show the same scaling behaviour. In addition, our results for the higher correlation functions $C_{a,1}$ and $C_{2,2}$ are also consistent with the asymptotic behaviours predicted by the $c=-2$ LCFT.

\paragraph{Square lattice amplitudes.}
For the two-point correlations $C_{1,1}$, $C_{2,1}$, $C_{3,1}$, $C_{4,1}$ (reconstructed from the sum rule), and $C_{2,2}$, the leading amplitudes $\mathcal{A}_{ab}$ agree with the exact predictions to within about $12\%$ (Table~\ref{tab:square_fits}). In particular, the reconstructed $C_{4,1}$ provides a non-trivial consistency check: its fitted amplitude $(8.519\pm0.203)\times 10^{-3}$ is positive, as required by the sum rule and the FSC mechanism, and matches the exact prediction $8.480\times 10^{-3}$ to within $0.5\%$. The $L=8000\,\ell$ data also reduce the effective finite-size ratio to $r/N\sim0.4\%$.

\paragraph{Honeycomb lattice.}
For the honeycomb lattice, the exact one-site probabilities $P_1=1/12$, $P_2=7/24$ and $P_3=5/8$ are reproduced to high precision (Table~\ref{tab:honeycomb_heights}). The fitted $C_{1,1}$ exponent from the sublattice-averaged curve is again close to $2\Delta=4$, and the fitted constant in $r^4 C_{1,1}(r)$ is within about $5.3\%$ of the exact honeycomb asymptotic coefficient $-9/(96\pi^2)$ in $\ell$-units (Table~\ref{tab:honeycomb_amplitudes}). The data for $C_{2,1}$ and $C_{3,1}$ confirm the expected linear dependence on $\log r$, while $C_{2,2}$ exhibits the expected quadratic dependence; the reported inter-sublattice standard deviations quantify the residual difference between the two honeycomb sublattices.

\paragraph{Kagome lattice.}
To our knowledge, this is the first systematic study of sandpile correlations on the kagome lattice. The key findings are:
\begin{itemize}
  \item The power-law exponent $2\Delta \approx 4$ is consistent with the same scaling law.
  \item The three sublattices have statistically identical height probabilities and correlation functions; this is reflected in the small standard deviations reported for the averaged fits.
  \item The functional form of the rescaled correlations---constant, linear, and quadratic in $\log r$ for Cases~1, 2, and~3, respectively (Table~\ref{tab:kagome_amplitudes})---matches the LCFT predictions of Section~\ref{sec:CFT}.
  \item In the common nearest-neighbour ($\ell$) convention, the kagome amplitudes are of the same order as their square-lattice counterparts. The inferred coefficients $\alpha_2=0.869\pm0.151$, $\alpha_3=1.980\pm0.085$, and $\alpha_4=-2.808\pm0.263$ are likewise comparable in magnitude to the exact square-lattice values (Table~\ref{tab:expansion parameters}).
\end{itemize}

\paragraph{Outlook.}
Several directions for future work present themselves: (i) extraction of the remaining scaling-field coefficients $\beta_a$ for the honeycomb and kagome from the numerical data, which requires either larger lattices or more samples to disentangle the asymptotic amplitudes from sub-leading corrections; (ii) extension to other lattice geometries like the triangular lattice; (iii) study of higher-point correlations and their comparison with LCFT predictions; (iv) analytical derivation of further bulk amplitudes on the honeycomb and kagome lattices, beyond the quantities currently known exactly.

%%%%%%%%%%%%%%%%%%%%%%%%%%%%
\section*{Acknowledgements}
%%%%%%%%%%%%%%%%%%%%%%%%%%%%
DS would like to thank the groups for mathematical physics and many-body theory at the University of Wuppertal for hospitality, and in particular Philippe Ruelle for numerous very valuable discussions. We also thank Jacopo Fiore for useful discussions related to the analytic derivation of the bulk height-1 probability on the kagome lattice. This work was funded by the Deutsche Forschungsgemeinschaft (DFG, German Research Foundation) -- 508440990. Simulations were performed with computing resources granted by RWTH Aachen University under projects rwth1807 and rwth1543.

%%%%%%%%%%%%%%%%%%%%%%%%%%%%
\section*{AI disclosure}
%%%%%%%%%%%%%%%%%%%%%%%%%%%%
Generative artificial intelligence (AI) tools (ChatGPT 5.4 by OpenAI) were used to assist with code development and to improve the language and presentation of the manuscript. In Section~\ref{sec:kagomeP1analytical}, AI was instructed to symbolically solve the required integrals using the SymPy python library and check against numerical results . The results were analytically verified by the authors, as documented in the derivations presented in Appendix A. The authors take full responsibility for the content and conclusions of this work.

%%%%%%%%%%%%%%%%%%%%%%%%%%%%
\appendix
\section{Derivation of the Green function values for the kagome lattice}
\label{app:kagome_green}
%%%%%%%%%%%%%%%%%%%%%%%%%%%%
Here, we collect the Brillouin-zone integral derivations of the relative Green-function values used in Sec.~5.3. These are the identities entering the construction of the local matrix $G_{\mathrm{loc}}$ for the height-1 defect problem. The real-space Green's function centered around a given site is obtained by Fourier transforming the inverse toppling matrix
\begin{equation}
G_{\alpha\beta}(n_1,n_2)
=
\frac{1}{4\pi^2}
\int_0^{2\pi}\!\! d\theta_1
\int_0^{2\pi}\!\! d\theta_2\,e^{-\ii(n_1\theta_1 + n_2\theta_2)}
\widetilde{\Delta}(\theta_1,\theta_2)^{-1}_{\alpha\beta}\;.
\end{equation}
In order for this integral to converge, a regularization is required, which is obtained by instead computing the integral of the difference $G - G_{AA}(0,0)$. This choice is ambiguous, since all sublattice components diverge with the same power law at $n_1=n_2=0$. In the following we will only compute those terms that cannot be related through bulk symmetries.

%%%%%%%%%%%%%%%%%%%%%%%%%%%%
\subsection{Derivation of $G_{AC}(0,0)-G_{AA}(0,0)$}
It is
\begin{equation}
G_{AC}(0,0)-G_{AA}(0,0)
=
\frac{1}{4\pi^2}
\int_0^{2\pi}\!\! d\theta_1
\int_0^{2\pi}\!\! d\theta_2\,
\Bigl[\widetilde{\Delta}(\theta_1,\theta_2)^{-1}_{AC}
- \widetilde{\Delta}(\theta_1,\theta_2)^{-1}_{AA}\Bigr]
\end{equation}
Defining the determinant attributed to the inverse toppling matrix as 
\begin{equation}
D(\theta_1,\theta_2)
=
3-\cos\theta_1-\cos\theta_2-\cos(\theta_1-\theta_2),
\end{equation}
the required matrix elements are
\begin{equation}
\widetilde{\Delta}(\theta_1,\theta_2)^{-1}_{AA}
=
\frac{7-\cos\theta_1}{6D},
\qquad
\widetilde{\Delta}(\theta_1,\theta_2)^{-1}_{AC}
=
\frac{5+e^{-\ii\theta_1}+e^{\ii\theta_2}+5e^{-\ii(\theta_1-\theta_2)}}{12D}.
\end{equation}
This yields for  the difference
\begin{align}
&\widetilde{\Delta}(\theta_1,\theta_2)^{-1}_{AC}
-\widetilde{\Delta}(\theta_1,\theta_2)^{-1}_{AA}\nonumber\\
&=
\frac{
3\cos\theta_1+\cos\theta_2+5\cos(\theta_1-\theta_2)-9
+ \ii\left[-\sin\theta_1+\sin\theta_2-5\sin(\theta_1-\theta_2)\right]
}{
12D(\theta_1,\theta_2)
}.
\end{align}
The imaginary part does not contribute to the integral due to the symmetric integration boundaries and we are left with the real part,
\begin{equation}
    R(\theta_1,\theta_2) = \mathrm{Re}\left[ \widetilde{\Delta}(\theta_1,\theta_2)^{-1}_{AC}-\widetilde{\Delta}(\theta_1,\theta_2)^{-1}_{AA}\right]\;.
\end{equation}
Its numerator can be rewritten as
\begin{equation}
3\cos\theta_1+\cos\theta_2+5\cos(\theta_1-\theta_2)-9
=
-3D(\theta_1,\theta_2)
+2\Bigl[\cos(\theta_1-\theta_2)-\cos\theta_2\Bigr],
\end{equation}
so that
\begin{equation}
R(\theta_1,\theta_2)
=
-\frac14
+\frac{\cos(\theta_1-\theta_2)-\cos\theta_2}{6D(\theta_1,\theta_2)}.
\end{equation}
Therefore,
\begin{equation}
G_{AC}(0,0)-G_{AA}(0,0)
=
-\frac14
+\frac{1}{24\pi^2}
\int_0^{2\pi}\!\! d\theta_1
\int_0^{2\pi}\!\! d\theta_2\,
\frac{\cos(\theta_1-\theta_2)-\cos\theta_2}{D(\theta_1,\theta_2)}.
\end{equation}
To show that the remaining integral vanishes, it is enough to prove that the first and second summand actually contribute equally to the integral. For that we perform the change of variables
$u=\theta_1,
\quad
v=\theta_1-\theta_2$,
which transforms first summand as
\begin{align}
\int_0^{2\pi}\!\! d\theta_1
\int_0^{2\pi}\!\! d\theta_2\,
\frac{\cos\theta_2}{D(\theta_1,\theta_2)} &\hspace{1.5mm}=
\int_0^{2\pi}\!\! du
\int_0^{2\pi}\!\! dv\,
\frac{\cos(u-v)}{3-\cos u-\cos(u-v)-\cos v} \nonumber\\
&\overset{\substack{u\equiv\theta_1 \\ v\equiv\theta_2}}{=} \int_0^{2\pi}\!\! d\theta_1
\int_0^{2\pi}\!\! d\theta_2\,
\frac{\cos(\theta_1-\theta_2)}{D(\theta_1,\theta_2)},
\end{align}
and therefore the second integral is indeed zero. We conclude that
\begin{equation}
\boxed{
G_{AC}(0,0)-G_{AA}(0,0)
=
-\frac14.}
\end{equation}

%%%%%%%%%%%%%%%%%%%%%%%%%%%%
\subsection{Derivation of $G_{BC}(-1,1)-G_{AA}(0,0)$}\label{app:second_derivation}
We now consider
\begin{equation}
G_{BC}(-1,1)-G_{AA}(0,0)
=
\frac{1}{4\pi^2}
\int_0^{2\pi}\!\! d\theta_1
\int_0^{2\pi}\!\! d\theta_2\,
\left[
e^{\ii(\theta_1-\theta_2)}\widetilde{\Delta}(\theta_1,\theta_2)^{-1}_{BC}
-
\widetilde{\Delta}(\theta_1,\theta_2)^{-1}_{AA}
\right].
\end{equation}
Using
\begin{equation}
\widetilde{\Delta}(\theta_1,\theta_2)^{-1}_{AA}
=
\frac{7-\cos\theta_1}{6D},
\qquad
\widetilde{\Delta}(\theta_1,\theta_2)^{-1}_{BC}
=
\frac{5+5e^{-\ii\theta_1}+e^{-\ii\theta_2}+e^{-\ii(\theta_1-\theta_2)}}{12D},
\end{equation}
we obtain
\begin{align}
G_{BC}(-1,1)&-G_{AA}(0,0)\nonumber\\
&\hspace{-10mm}=
\frac{1}{4\pi^2}
\int_0^{2\pi}\!\! d\theta_1
\int_0^{2\pi}\!\! d\theta_2\,
\frac{
2\cos\theta_1+5\cos\theta_2+\cos(\theta_1-2\theta_2)+5\cos(\theta_1-\theta_2)-13
}{
12D(\theta_1,\theta_2)
},
\end{align}
since the imaginary part again vanishes upon integration by symmetry. To reduce this to a one-angle integral, we use
\begin{equation}
\theta_2=2\phi,
\qquad
\alpha=\theta_1-\phi,\quad\text{
so that}\quad
\theta_1=\alpha+\phi,
\qquad
\theta_1-\theta_2=\alpha-\phi.
\end{equation}
Using the shorthands
\begin{equation}
c=\cos\phi,
\qquad
s=\sin\phi,
\end{equation}
the denominator can be written as 
\begin{equation}
D=2\bigl(1+\sin^2\phi-\cos\phi\cos\alpha\bigr)
=
2(1+s^2-c\cos\alpha),
\end{equation}
while the numerator turns into
\begin{align}
&2\cos(\alpha+\phi)+5\cos 2\phi+\cos(\alpha-3\phi)+5\cos(\alpha-\phi)-13
\nonumber\\
&\qquad\qquad=
2s(1+2c^2)\sin\alpha+4c(1+c^2)\cos\alpha+10c^2-18.
\end{align}
The term proportional to $\sin\alpha$ is odd in $\alpha$ and integrates to zero. We can now integrate over $\alpha$ using the standard integrals
\begin{equation}\label{eq:standard_integrals}
\int_0^{2\pi}\frac{d\alpha}{a-b\cos\alpha}
=
\frac{2\pi}{\sqrt{a^2-b^2}},\qquad
\int_0^{2\pi}\frac{\cos\alpha\,d\alpha}{a-b\cos\alpha}
=
\frac{1}{b}\left(\frac{2\pi a}{\sqrt{a^2-b^2}}-2\pi\right),
\end{equation}
with
\begin{equation}
a=1+s^2,
\qquad
b=c.
\end{equation}
One obtains
\begin{equation}
G_{BC}(-1,1)-G_{AA}(0,0)
=
-\frac{1}{12\pi}\int_0^\pi
\left[
2(1+c^2)
+
\frac{s\,(5-2c^2)}{\sqrt{4-c^2}}
\right]d\phi.
\end{equation}
To solve the remaining integral, we change the integration variable by setting
\begin{equation}
d\phi=-\frac{dc}{\sqrt{1-c^2}} = -\frac{dc}{s},
\end{equation}
to obtain
\begin{equation}
G_{BC}(-1,1)-G_{AA}(0,0)
=
-\frac{1}{12\pi}\int_{-1}^{1}
\left[
\frac{2(1+c^2)}{\sqrt{1-c^2}}
+
\frac{5-2c^2}{\sqrt{4-c^2}}
\right]dc.
\end{equation}
The two contributions are elementary,
\begin{equation}
\int_{-1}^{1}\frac{2(1+c^2)}{\sqrt{1-c^2}}\,dc
=3\pi,
\end{equation}
and, by setting $c=2\sin t$,
\begin{equation}
\int_{-1}^{1}\frac{5-2c^2}{\sqrt{4-c^2}}\,dc
=
\int_{-\pi/6}^{\pi/6}(5-8\sin^2 t)\,dt
=
\frac{\pi}{3}+2\sqrt3.
\end{equation}
Hence, the final result becomes
\begin{equation}
\boxed{
G_{BC}(-1,1)-G_{AA}(0,0)
=
-\frac{5}{18}-\frac{\sqrt3}{6\pi}.}
\end{equation}

%%%%%%%%%%%%%%%%%%%%%%%%%%%%
\subsection{Derivation of $G_{CC}(-1,1)-G_{AA}(0,0)$}
Finally, consider
\begin{equation}
G_{CC}(-1,1)-G_{AA}(0,0)
=
\frac{1}{4\pi^2}
\int_0^{2\pi}\!\! d\theta_1
\int_0^{2\pi}\!\! d\theta_2\,
\left[
e^{\ii(\theta_1-\theta_2)}\widetilde{\Delta}(\theta_1,\theta_2)^{-1}_{CC}
-
\widetilde{\Delta}(\theta_1,\theta_2)^{-1}_{CC}
\right].
\end{equation}
Here we used the bulk sublattice symmetry $G_{CC}(0,0)=G_{AA}(0,0)$. Since
\begin{equation}
\widetilde{\Delta}(\theta_1,\theta_2)^{-1}_{CC}
=
\frac{7-\cos\theta_2}{6D(\theta_1,\theta_2)},
\end{equation}
we have
\begin{equation}
G_{CC}(-1,1)-G_{AA}(0,0)
=
\frac{1}{4\pi^2}
\int_0^{2\pi}\!\! d\theta_1
\int_0^{2\pi}\!\! d\theta_2\,
\frac{(7-\cos\theta_2)\bigl(e^{\ii(\theta_1-\theta_2)}-1\bigr)}{6D(\theta_1,\theta_2)}.
\end{equation}
The imaginary part again vanishes upon integration, so we may write
\begin{align}
G_{CC}(-1,1)&-G_{AA}(0,0) \nonumber\\
&\hspace{-10mm}=
\frac{1}{4\pi^2}
\int_0^{2\pi}\!\! d\theta_1
\int_0^{2\pi}\!\! d\theta_2\,
\frac{
14\cos(\theta_1-\theta_2)-2\cos\theta_2\cos(\theta_1-\theta_2)+2\cos\theta_2-14
}{
12D(\theta_1,\theta_2)
}.
\end{align}
We now perform the same one-angle reduction as in Appendix~\ref{app:second_derivation}. The denominator becomes
\begin{equation}
D=2\bigl(1+\sin^2\phi-\cos\phi\cos\alpha\bigr)
=
2(1+s^2-c\cos\alpha),
\end{equation}
while the numerator simplifies to
\begin{align}
&14\cos(\alpha-\phi)-2\cos 2\phi\,\cos(\alpha-\phi)+2\cos 2\phi-14
\nonumber\\
&\qquad=
4s(4-c^2)\sin\alpha+4c(4-c^2)\cos\alpha+4(c^2-4).
\end{align}
The term proportional to $\sin\alpha$ is odd and therefore integrates to zero. Using again the standard integrals in Eq.~\eqref{eq:standard_integrals}, the $\alpha$ integral is solved as
\begin{align}
G_{CC}(-1,1)-G_{AA}(0,0)
&=
\frac{1}{6\pi}\int_0^\pi
\bigl[
s\sqrt{4-c^2}-(4-c^2)
\bigr]d\phi
\nonumber\\
&=
\frac{1}{6\pi}\int_{-1}^{1}
\left[
\sqrt{4-c^2}
+
\frac{c^2-4}{\sqrt{1-c^2}}
\right]dc.
\end{align}
The two contributions are elementary,
\begin{equation}
\int_{-1}^{1}\frac{c^2-4}{\sqrt{1-c^2}}\,dc
=
\int_{-1}^{1}\frac{c^2\,dc}{\sqrt{1-c^2}}
-4\int_{-1}^{1}\frac{dc}{\sqrt{1-c^2}}
=
\frac{\pi}{2}-4\pi=-\frac{7\pi}{2},
\end{equation}
while, by setting $c=2\sin t$,
\begin{equation}
\int_{-1}^{1}\sqrt{4-c^2}\,dc
=
\int_{-\pi/6}^{\pi/6}4\cos^2 t\,dt
=
\frac{2\pi}{3}+\sqrt3.
\end{equation}
Therefore
\begin{equation}
\boxed{
G_{CC}(-1,1)-G_{AA}(0,0)
=
\frac{1}{6\pi}\left(-\frac{7\pi}{2}+\frac{2\pi}{3}+\sqrt3\right)
=
-\frac{17}{36}+\frac{\sqrt3}{6\pi}.
}
\end{equation}

%%%%%%%%%%%%%%%%%%%%%%%%%%%%
%The bibliographystyle commands should be commented out, since they are apparently included in the SciPost class file.


\begin{thebibliography}{1}
\providecommand{\url}[1]{\texttt{#1}}
\providecommand{\urlprefix}{URL }
\expandafter\ifx\csname urlstyle\endcsname\relax
  \providecommand{\doi}[1]{doi:\discretionary{}{}{}#1}\else
  \providecommand{\doi}{doi:\discretionary{}{}{}\begingroup
  \urlstyle{rm}\Url}\fi
\providecommand{\eprint}[2][]{\url{#2}}

\bibitem{Bak-87}
P.~Bak, C.~Tang and K.~Wiesenfeld,
\newblock \emph{Self-organized criticality: An explanation of the 1/f noise},
\newblock Phys. Rev. Lett. \textbf{59}, 381 (1987),
\newblock \doi{10.1103/PhysRevLett.59.381}.

\bibitem{Bak-88}
P.~Bak, C.~Tang and K.~Wiesenfeld,
\newblock \emph{Self-organized criticality},
\newblock Phys. Rev. A \textbf{38}, 364 (1988),
\newblock \doi{10.1103/PhysRevA.38.364}.

\bibitem{Dhar90}
D.~Dhar,
\newblock \emph{Self-organized critical state of sandpile automaton models},
\newblock Phys. Rev. Lett. \textbf{64}, 1613 (1990),
\newblock \doi{10.1103/PhysRevLett.64.1613},
\newblock Erratum: \emph{ibid.} \textbf{64}, 2837 (1990).

\bibitem{Dhar06}
D.~Dhar,
\newblock \emph{Theoretical studies of self-organized criticality},
\newblock Physica A \textbf{369}, 29 (2006),
\newblock \doi{10.1016/j.physa.2006.04.004}.

\bibitem{BPZ84}
A.~A. Belavin, A.~M. Polyakov and A.~B. Zamolodchikov,
\newblock \emph{Infinite conformal symmetry in two-dimensional quantum field theory},
\newblock Nucl. Phys. B \textbf{241}, 333 (1984),
\newblock \doi{10.1016/0550-3213(84)90052-X}.

\bibitem{DiFrancescoMathieuSenechal97}
P.~Di Francesco, P.~Mathieu D.~S\'{e}n\'{e}chal,
\newblock \emph{{C}onformal {F}ield {T}heory},
\newblock (Springer, New York, 1997).

\bibitem{Henkel99}
M.~Henkel,
\newblock \emph{Conformal invariance and critical phenomena},
\newblock (Springer, Berlin, 1999).

\bibitem{GRR13}
A.~Gainutdinov, D.~Ridout and I.~Runkel (eds.),
\newblock \emph{Logarithmic conformal field theories},
\newblock special issue of J. Phys. A: Math. Theor. \textbf{46} (2013),
\newblock \doi{10.1088/1751-8113/46/49/490301}.

\bibitem{Ruelle21}
P.~Ruelle,
\newblock \emph{Sandpile models in the large},
\newblock Frontiers in Physics \textbf{9} (2021),
\newblock \doi{10.3389/fphy.2021.641966},
\newblock Topic: Self-Organized Criticality, Three Decades Later.

\bibitem{MajumdarDhar92}
S.~N. Majumdar and D.~Dhar,
\newblock \emph{Equivalence between the Abelian sandpile model
  and the $q \to 0$ limit of the Potts model},
\newblock Physica A \textbf{185}, 129 (1992),
\newblock \doi{10.1016/0378-4371(92)90016-I}.

\bibitem{PGPR08}
V.~S. Poghosyan, S.~Y. Grigorev, V.~B. Priezzhev and P.~Ruelle,
\newblock \emph{Pair correlations in sandpile model: A check of logarithmic conformal field theory},
\newblock Phys. Lett. B \textbf{659}, 768 (2008),
\newblock \doi{10.1016/j.physletb.2007.12.002}.

\bibitem{PGPR10}
V.~S. Poghosyan, S.~Y. Grigorev, V.~B. Priezzhev and P.~Ruelle,
\newblock \emph{Logarithmic two-point correlators in the Abelian sandpile model},
\newblock J. Stat. Mech.: Theory Exp. \textbf{2010}, P07025 (2010),
\newblock \doi{10.1088/1742-5468/2010/07/P07025}.

\bibitem{PonceletRuelle17}
A.~Poncelet and P.~Ruelle,
\newblock \emph{Multipoint correlators in the Abelian sandpile model},
\newblock J. Stat. Mech.: Theory Exp. \textbf{2017}, 123102 (2017),
\newblock \doi{10.1088/1742-5468/aa9bdb}.

\bibitem{JPR06}
M.~Jeng, G.~Piroux and P.~Ruelle,
\newblock \emph{Height variables in the Abelian sandpile model: Scaling fields and correlations},
\newblock J. Stat. Mech.: Theory Exp. \textbf{2006}, P10015 (2006),
\newblock \doi{10.1088/1742-5468/2006/10/P10015}.

\bibitem{PR05}
G.~Piroux and P.~Ruelle,
\newblock \emph{Logarithmic scaling for height variables in the Abelian sandpile model},
\newblock Phys. Lett. B \textbf{607}, 188 (2005),
\newblock \doi{10.1016/j.physletb.2004.12.045}.

\bibitem{Wilson96}
D.~B. Wilson,
\newblock \emph{Generating random spanning trees more quickly than the cover time},
\newblock In \emph{Proceedings of the Twenty-Eighth Annual ACM Symposium on Theory of Computing (STOC '96)}, pp.~296--303 (1996),
\newblock \doi{10.1145/237814.237880}.

\bibitem{KytolaRidout09}
K.~Kyt{\"o}l{\"a} and D.~Ridout,
\newblock \emph{On staggered indecomposable Virasoso modules},
\newblock J. Math. Phys. \textbf{50}, 123503 (2009),
\newblock \doi{10.1063/1.3191682}.

\bibitem{MajumdarDhar91}
S.~N. Majumdar and D.~Dhar,
\newblock \emph{Height correlations in the Abelian sandpile model},
\newblock J. Phys. A: Math. Gen. \textbf{24}, L357 (1991),
\newblock \doi{10.1088/0305-4470/24/7/008}.

\bibitem{PonceletRuelle18}
A.~Poncelet and P.~Ruelle,
\newblock \emph{Sandpile probabilities on triangular and hexagonal lattices},
\newblock J. Phys. A: Math. Theor. \textbf{51}, 095003 (2018),
\newblock \doi{10.1088/1751-8121/aa9255}.

\bibitem{Honeycomb10}
N.~Azimi-Tafreshi, H.~Dashti-Naserabadi, S.~Moghimi-Araghi and P.~Ruelle,
\newblock \emph{The Abelian sandpile model on the honeycomb lattice},
\newblock J. Stat. Mech.: Theory Exp. \textbf{2010}, P02004 (2010),
\newblock \doi{10.1088/1742-5468/2010/02/P02004}.

\bibitem{KasselWilson16}
A.~Kassel and D.~B. Wilson,
\newblock \emph{The looping rate and sandpile density of planar graphs},
\newblock The American Mathematical Monthly \textbf{123}, 19 (2016),
\newblock \doi{10.4169/amer.math.monthly.123.1.19}.

\end{thebibliography}
\end{document}